\documentclass[prb,reprint,amsmath,amssymb,showpacs,superscriptaddress,floatfix,longbibliography]{revtex4-1}
\usepackage[breaklinks=true,colorlinks,citecolor=blue,linkcolor=blue,urlcolor=blue]{hyperref}

\usepackage{epsfig,mathrsfs,color,latexsym,subfigure,marginnote,gensymb}
\renewcommand{\BibitemShut}[1]{}

\begin{document}
\title{Four allotropes of semiconducting layered Arsenic which switch into a topological insulator via an electric field: A computational study}

\author{Sougata Mardanya}
\affiliation{Department of Physics, Indian Institute of Technology, Kanpur, Kanpur 208016, India}
\author{Vinay Kumar Thakur}
\affiliation{Dept. of Material Science and Engineering, Indian Institute of Technology, Kanpur, Kanpur 208016, India}
\author{Somnath Bhowmick}
\email[]{bsomnath@iitk.ac.in}
\affiliation{Dept. of Material Science and Engineering, Indian Institute of Technology, Kanpur, Kanpur 208016, India}
\author{Amit Agarwal}
\email{amitag@iitk.ac.in}
\affiliation{Department of Physics, Indian Institute of Technology, Kanpur, Kanpur 208016, India}

\date{\today}
\begin{abstract}
We propose four different thermodynamically stable structural phases of arsenic monolayers based on {\it ab-initio} density functional theory calculations all of which undergo a topological phase transition on application of a perpendicular electric field. All the four arsenic monolayer allotropes have a wide band gap, varying from 1.21 eV to 3.0 eV (based on GW calculations),  and in general they undergo a metal-insulator quantum phase transition on application  of uniaxial in-layer strain.  Additionally an increasing  transverse electric field induces band-inversion at the $\Gamma$ point in all four monolayer allotropes, leading to a nontrivial topological phase (insulating for three and metallic for one allotrope), characterized by the switching of the $Z_2$ index, from 0 (before band inversion) to 1 (after band inversion). The topological phase tuned by the transverse electric field, should support spin-separated gapless edge states which should manifest in quantum spin Hall effect.
\end{abstract}
\maketitle
\section{Introduction}
Since the discovery of graphene,\cite{graphene1, RevModPhys.81.109, 9781139031080} various other 2D crystals such as, silicine,\cite{PhysRevB.50.14916, PhysRevB.76.075131, 1.3524215, silicene2, PhysRevLett.109.056804} germanene,\cite{PhysRevB.76.075131, germanane1,PhysRevLett.102.236804} transition metal dichalcogenides (MoS$_2$, MoSe$_2$, WSe$_2$ etc.),\cite{TMD1,10.1038/nphys2942, MoS2_priyank} and phosphorene,\cite{BP-transistor, phosphorene1, APL_Neto} are being actively explored for applications in nano electronics, optoelectronics and flexible electronics. For example, graphene,\cite{graphene_tr} MoS$_2$\cite{MoS2_tr} and black phosphorene\cite{BP-transistor, phosphorene1, APL_Neto} based field effect transistors have been fabricated successfully. Among these, graphene\cite{graphene1, RevModPhys.81.109, 9781139031080} has a very high mobility but is limited by its lack of an intrinsic band gap, and transition metal dichalcogenides (MoS$_2$, MoSe$_2$, WSe$_2$ etc.) \cite{TMD1,10.1038/nphys2942} while having a direct and large band gap, are limited in their mobility. As a consequence layered structures of isoelectronic systems, such as phosphorene, look very promising and are being actively explored on account of their high mobility combined with robust intrinsic band gap, which can be tuned by means of strain, transverse electric field, number of layers, adatoms etc. \cite{phosphorene1, BP-Qiao,  PRLNeto, zhu14, PRL.113.046804, BP-Ezawa, Ghosh1, Priyank_Padatom, Suhas_gs} More recently, new 2D semiconductors based on few layers of arsenic, with a puckered (similar to black phosphorous) and buckled  (similar to black phosphorous) honeycomb structure have been predicted,\cite{Zhu_As,Kamal} which are expected to be less reactive and have higher mobility than compared to phosphorene. Both of these mono-layered allotropes of arsenic are being actively studied in various contexts.\cite{Zerati, CaoSM, Zhiya, Shengli_APL, Yaping_APE, ZhenZhu_NL, Zhang_AIPA, Han_APE, Kou_JPCC}

{Motivated by Ref.~[\onlinecite{PRL.113.046804}], wherein two new forms of layered phosphorene were discussed, in this article we propose two additional stable (mono) layered structures of arsenic, other than the already predicted puckered and buckled form. Furthermore, we study the stability, structural and electronic properties of all four monolayer allotropes, including the effect of a transverse electric field and uniaxial in-layer strain.} The nomenclature of $\alpha-$ (puckered honeycomb structure analogous to black phosphorene), $\beta$- (buckled honeycomb structure analogous to blue phosphorous,\cite{Ghosh1} and known as grey arsenene), $\gamma$- and $\delta$-arsenic  is chosen, keeping in mind their structural similarities with monolayer phosphorous allotropes.\cite{PRL.113.046804} We find that, other than $\gamma$, rest of the allotropes are indirect band gap semiconductors. Interestingly, all of them undergo band inversion around the time reversal invariant $\Gamma$-point in the Brillouin zone, beyond a certain critical electric field applied in the transverse direction. Moreover, we also find that the band inversion is accompanied by a topological phase transition, as confirmed by the calculation of the $Z_2$ index \cite{z2kane, z2Rahul2, z2Rahul} and the topological phase should support spin-separated gap-less edge states which can give rise to quantum spin Hall effect. A metal-insulator transition is also observed under the influence of uniaxial strain, applied in-plane of the monolayers. We also have confirmed the stability of the allotropes at room temperature and beyond, by doing \textit{ab initio} molecular dynamics (MD) simulations.

The article is organized as follows: In Sec.~\ref{SecII} we briefly describe the details of the density-functional theory calculations, which is followed by a discussion of the structural and electronic properties of all the four allotropes, and their stability based on molecular dynamics simulation. In Sec.~\ref{SecIV}, we study the effect of a transverse electric field on the band structure of all the four allotropes, and explicitly demonstrate the phenomena of  band-inversion and topological phase transition in all the four monolayer allotropes with increasing electric field. This is followed by a study of the in-plane uniaxial strain on the electronic properties in Sec.~\ref{SecV}, and finally we summarize our findings in Sec.~\ref{SecVI}. 
\begin{figure*}
\begin{center}
\includegraphics[width=0.99 \linewidth]{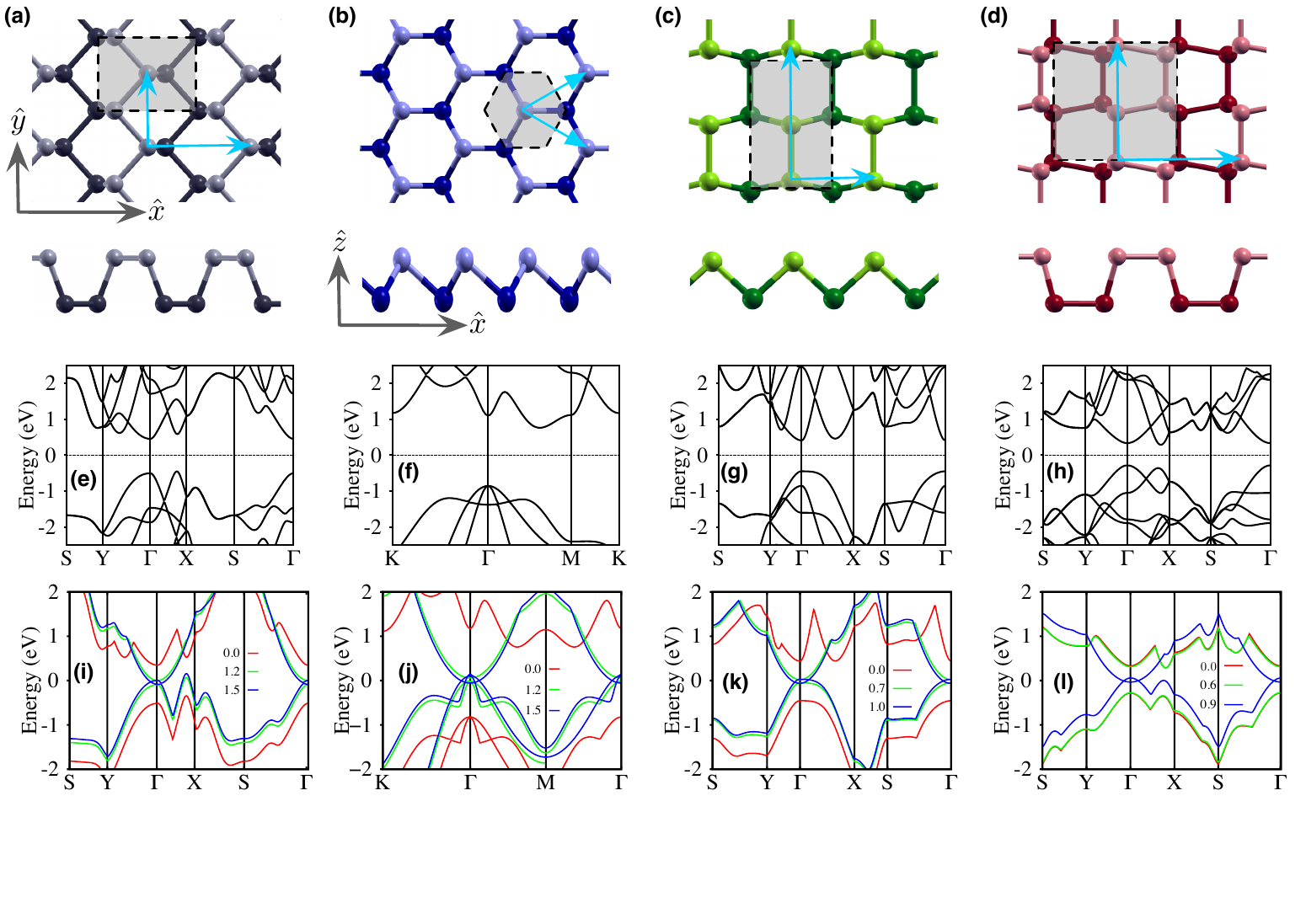}
\caption{ Panels (a), (b), (c) and (d) display the equilibrium structure (top and side view), panels (e), (f), (g) and (h) display the zero field intrinsic band structure along high-symmetry directions and  finally panels (i), (j), (k) and (l) highlight the valence and conduction band of $\alpha$-As (puckered), $\beta$-As (grey or buckled), $\gamma$-As and and $\delta$-As, respectively, at finite electric field applied perpendicular to the monolayer arsenic allotropes (for three different electric field strengths).  The shaded region in panels (a)-(d) depicts the Wigner-Seitz cell and the blue arrows mark the lattice vectors. The atoms belonging to the top and bottom layer of non-planar structures are shown in light and dark shade, respectively. The electric-field induced band inversion, which is also accompanied by a topological phase,  in all four allotropes with increasing electric field is evident in panels (i)-(l).} 
\label{f1}
\end{center}
\end{figure*}

\section{lattice structure, stability and band structure }

 \label{SecII}
The equilibrium structure, stability and electronic properties are analyzed using the {\it ab initio} density functional theory (DFT) based calculations, using a plane-wave basis set and ultrasoft pseudopotentials, as implemented in Quantum Espresso \cite{QE}.  Electron exchange and correlation is treated within the framework of generalized gradient approximation (GGA) given by Perdew-Burke-Ernzerhof  functional \cite{PBE}. Since GGA-PBE exchange-correlation functional underestimates the bandgap, we also calculate the electronic band structure using the well-known GW approximation \cite{Hedin, Kresse}, which has been shown to reproduce the experimentally measured bandgap in monolayer black phosphorene \cite{Rudenko} and other materials. The kinetic energy cutoff for the wave function is set to be 40 Ry. Supercell with a vacuum of 25~\AA~ along the $z$ direction (perpendicular to the monolayers) is constructed to eliminate the interaction with spurious replica images. Brillouin zone integrations are performed using a Monkhorst-Pack scheme with a $k$-point grid of $24\times 24\times 1$ for $\beta$-arsenic. Since the step size of the $k$-point grid is inversely proportional to the length of the repeat vector in a given direction, we scale the grid size accordingly for other allotropes. Structural relaxations are carried out until the force on each atom (total energy change due to ionic relaxation between two successive steps) is less than $10^{-3}$ Ry/au ($10^{-4}$ Ry).

\begin{figure*}[t]
\begin{center}
\includegraphics[width=0.99 \linewidth]{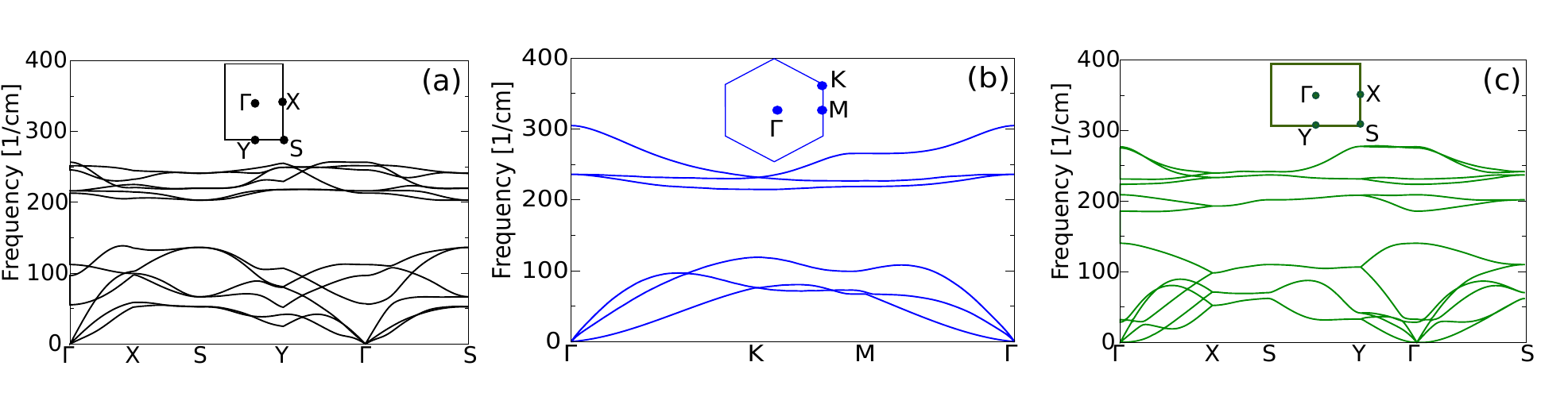}
\caption{Phonon dispersion of (a) $\alpha$-As, (b) $\beta$-As and (c) $\gamma$-As.}
\label{f1a}
\end{center}
\end{figure*}

Atomic arrangement and Wigner-Seitz cell of each of the arsenic allotropes is shown in Fig.~\ref{f1} and optimized structural parameters and cohesive energies ($E_{\rm c}$) are reported in Table~\ref{t1}. As illustrated in the diagram, $\beta$-arsenic has a highly symmetrical hexagonal unit cell and the rest of the allotropes have rectangular unit cell of lesser symmetry. Although $\beta$-arsenic has some resemblance to graphene's honeycomb structure, nevertheless, the constituent atoms does not form a flat 2D-crystal like graphene. Since As has one more electron in the $p$-orbital than compared to C, the planar honeycomb structure formed by $sp^2$ hybridization is not stable and $\beta$-arsenic has a buckled structure with non-zero thickness. Not only just $\beta$, monolayers of $\alpha$, $\gamma$ and $\delta$-arsenic also have some finite thickness because of the non-planar structure (see Table~\ref{t1}). Interestingly, bond length of triply coordinated As-atoms (see top view of the allotropes in Fig.~\ref{f1}) are found to be $\sim 2.5$ \AA~ for $\alpha$, $\beta$ and $\delta$-arsenic. Only $\gamma$-arsenic has one small exception; although the out of plane bonds are $\sim 2.5$~\AA~ long, in plane As-As bond is slightly larger ($2.57$ \AA) than compared to the equilibrium bond length [see Table~\ref{t1}]. Since the interatomic distances are close to their equilibrium value in all the allotropes, it is not surprising that all four allotropes have  similar cohesive energies ($E_{\rm c}$). Relatively small difference of  $E_{\rm c}$ (within 0.08 eV/atom) among the allotropes results from the bond angle variation, ranging from $92^{\degree}$ to $99.5^{\degree}$ [see Table~\ref{t1}]. We find that, among the four variants $\beta$-arsenic  has the highest $E_{\rm c}$ and is consequently the most stable, followed by $\alpha$, $\gamma$ and $\delta$-arsenic [see Table~\ref{t1}]. This is unlike phosphorous allotropes, where $\alpha$-phase is found to be the most stable one \cite{PRL.113.046804}. Note that the bond-lengths, bond-angle, monolayer thickness, and the cohesive energies reported by us are consistent (within reasonable bounds of $< 5\%$) with those reported for monolayer $\alpha$-As in Ref.~[\onlinecite{Kamal}], and for $\beta$-As in Refs.~[\onlinecite{Kamal, Zhu_As}].

\begin{figure*}[t]
\begin{center}
\includegraphics[width=0.99 \linewidth]{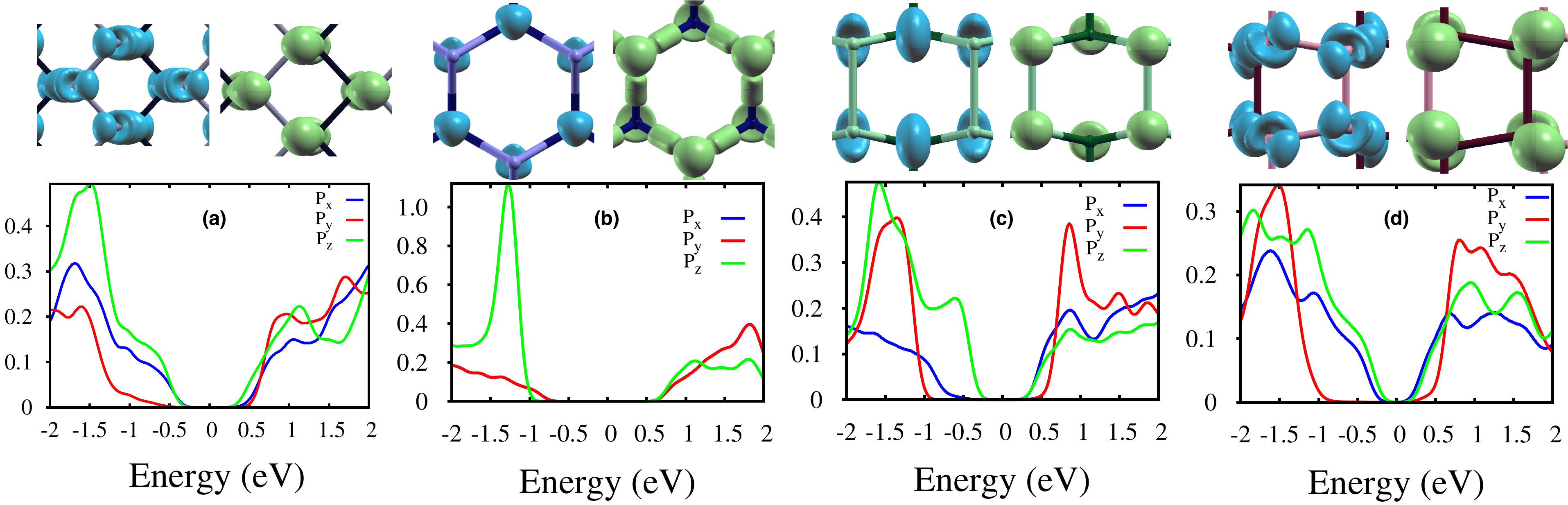}
\caption{Top view of the electron charge density near the valence (shaded in blue) and conduction (shaded in green) band edge is shown in the top panel and the orbital projected density of states (PDOS) is plotted in the bottom panel for all the four allotropes, a) $\alpha$-As, b) $\beta$-As, c) $\gamma$-As and d) $\delta$-As. The charge density is calculated for only the states within $\sim 0.2$  eV of the valence and conduction band edge.
\label{f2}}
\end{center}
\end{figure*}
\begin{table}[t]
\caption{Calculated values of equilibrium structural parameters, bandgap ($\Delta$) and cohesive energy ($E_{\rm c}$) for arsenic allotropes. Cohesive energy is calculated with respect to the total energy of an isolated As atom ($E_{\rm As}$), given by $E_{\rm c}=E_{\alpha/\beta/\gamma/\delta}-E_{\rm As}$, where $E_{\alpha/\beta/\gamma/\delta}$ is the energy per atom of the corresponding monolayer allotrope of  arsenic. Based on the $E_{\rm c}$ values, we find $\beta$-arsenic (monolayer grey arsenic) to be the most stable among the four allotropes. \label{t1}}
\centering
\begin{tabular}{c c c c c}
\hline
Phase:  & $\alpha$-As & $\beta$-As & $\gamma$-As & $\delta$-As \\
\hline
Bond lengths (\AA)& 2.50 & 2.50 & 2.57,2.50 & 2.50 \\
Bond angles (degree) & 94.5, 100.2 & 91.7 & 98.7, 91.7 & 99.8, 99.1 \\
Thickness (\AA) & 2.39 & 1.39 & 1.68 & 2.40 \\
$|\bf a_1| $ (\AA) & 4.82 & 3.61 & 3.58 & 5.91 \\
$|\bf a_2|$ (\AA) & 3.68 & 3.61 & 5.92 & 5.93 \\
$\Delta^b$ (eV)& 0.70 & 1.62 & 0.86$^a$ & 0.57 \\
$\Delta^c$ (eV)& 1.30 & 3.00 & 2.01$^a$ & 1.21\\
$E_{\rm c}$ (eV/atom)& 3.13 & 3.15 & 3.07 & 3.06 \\ 
\hline

\hline
\end{tabular} \\
{\footnotesize ~  $^a$ Direct bandgap; $^b$ GGA-PBE; $^c$ GW}
\end{table}

{Before proceeding further, we establish the stability of all four arsenic allotropes by means of first principles molecular dynamics (MD) simulations at finite temperature \cite{Car85}, and by calculating the phonon dispersions. As shown in Fig~\ref{f1a}, the phonon frequencies are found to be positive for $\alpha$, $\beta$ and $\gamma$-As, indicating the stability of these 2D layers in freestanding form. However, we observe large negative frequencies near the $\Gamma$ point for the out-of-plane acoustic mode of $\delta$-As, which indicates that $\delta-$As monolayer allotrope can only be stabilized on suitable substrate, that can damp the out-of-plane vibrations. MD calculations are performed by taking 4$\times$4 super cell for $\alpha, \beta$ and $\gamma$-arsenic and 3$\times$3 super cell for $\delta$-arsenic, while the $\Gamma$ point is used for Brillouin zone sampling. The time-step for MD simulation is taken to be 0.145 fs with a total run time of $\approx 1$ps. The fictitious electron mass is chosen to be 200 a.u to ensure that the electrons remain close to the ground state and good control of the conserved quantities are maintained. No spontaneous disintegration is observed when the $\alpha$-As, $\beta$-As and $\gamma$-As structures are equilibrated at 700 K, while $\delta$-As is found to be stable upto 500 K. Thus, we anticipate that devices made from any of the arsenic allotropes are going to be structurally stable at room temperature and well beyond that.}

Having established the stability of all four monolayer-As allotropes, we now focus on their electronic bandstructure. According to DFT-PBE based calculations, all four allotropes of monolayer arsenic are semiconductors and the GGA bandgap $\Delta$ varies from a minimum of 0.57 eV in $\delta$-arsenic to a maximum of 1.62 eV in $\beta$-arsenic. Additionally the GW calculations predict significantly larger values of $\Delta$ while maintaining the same trend as that of the DFT-PBE calculations  [see Table~\ref{t1}], as expected. Earlier works on monolayer $\alpha$-As, have reported the GGA (or PBE) bandgap to be $0.8$ eV\cite{Kamal}, $0.9$ eV\cite{Zhiya, ZhenZhu_NL, Han_APE}, and HSE06 bandgap to be $1.54$ eV \cite{Zhiya} which are comparable to our GGA value of $0.7$ eV and GW bandgap of $1.30$ eV. For monolayer $\beta$-As, the GGA  bandgap has been reported to be 1.6 eV \cite{Kamal, ZhenZhu_NL}, 1.71 eV \cite{Kamal}, 1.67 eV \cite{CaoSM}, 1.87 eV \cite{Shengli_APL} and 1.5 eV \cite{Kou_JPCC}, which is comparable to the value we find 1.62 eV.However for $\beta$-As, the HSE06 bandgap has been reported to be 2 eV \cite{Zhu_As} and 2.2 eV \cite{CaoSM} while our GW calculations indicate a much higher value of 3 eV for the same.  

The DFT-PBE band structures of all the four allotropes are presented in Fig.~\ref{f1}(e)-(h). Note that while the DFT-PBE method underestimates the bandgap, it generally predicts all the qualitative features correctly. As shown in Fig.~\ref{f1}(e)-(h), other than the $\gamma$-arsenic, rest of the allotropes are indirect bandgap semiconductors in zero electric field, although the magnitude of direct and indirect bandgap is nearly equal in case of $\alpha$, $\gamma$ and $\delta$-arsenic [see Fig.~\ref{f1}(e)-(h)]. This is due to the fact that, either the valence band or the conduction band or both of them are very close (less than 0.1 eV) to the valence band maximum (VBM) and conduction band minimum (CBM) in more than one locations in the Brillouin zone, at the $\Gamma$ point and at a point close to the X point in $\Gamma$X direction [see Fig.~\ref{f1}(e), (g) and (h)]. In case of $\beta$-arsenic, the VBM and CBM is located at the $\Gamma$ point and at a location close to the M point in $\Gamma$M direction, respectively [see Fig.~\ref{f1}(f)]. The direct bandgap at the $\Gamma$  point is significantly larger (0.34 eV) in magnitude than that of the indirect bandgap for $\beta$-As. 

To gain further insight into the orbitals contributing to the CBM and VBM, we calculate the orbital projected density of states (PDOS) which reveals that the states near the valence and conduction band edge to have the character of \textit{p} states [see bottom panel of Fig.~\ref{f2}]. Note the significant contribution from the $p_z$ states in the PDOS plots, having out of plane electron density perpendicular to the arsenic layers. This is further corroborated by the charge density corresponding to the states within $\sim 0.2$  eV of the valence and conduction band edge, shown in the top panel of Fig.~\ref{f2}. In case of multilayer structures, these states are going to overlap between the adjacent layers (known as interlayer hopping), resulting a band dispersion perpendicular to the plane. Because of this, multilayer arsenic is expected to have smaller bandgap than compared to it's respective monolayer structure. Similar effect is observed in phosphorous monolayers, where the bandgap is found to be inversely proportional to the number of layer for $\alpha, \beta$ and $\delta$-P, while $\gamma$-P becomes metallic for two layers and beyond.~\cite{PRL.113.046804}.

\begin{figure}
\begin{center}
\includegraphics[width=0.85 \linewidth]{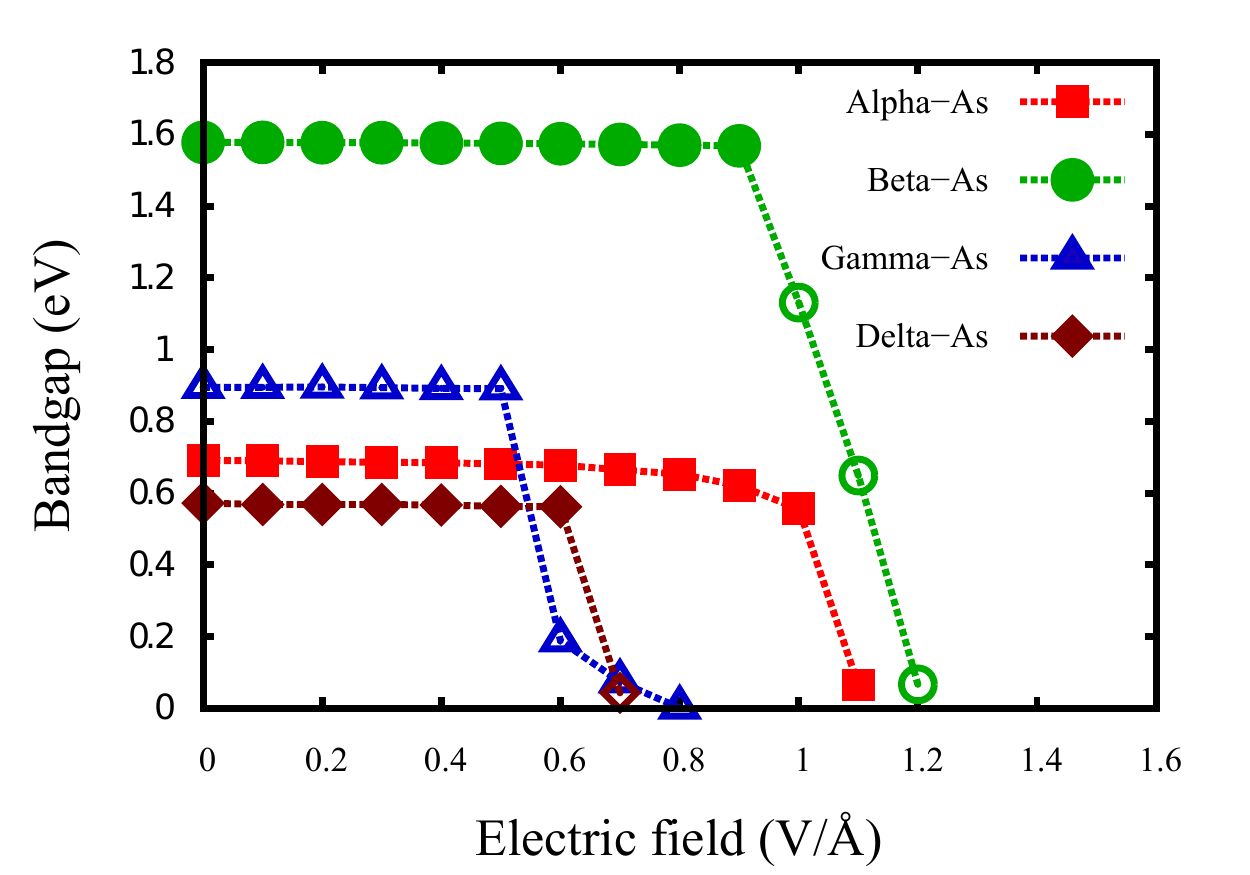}
\caption{Magnitude of bandgap (DFT-PBE) in monolayer arsenic allotropes is plotted as a function of vertically applied electric field. The open (filled) circles represent a direct (indirect) bandgap semiconductor.
\label{f3}}
\end{center}
\end{figure}

\begin{figure*}
\begin{center}
\includegraphics[width=0.85 \linewidth]{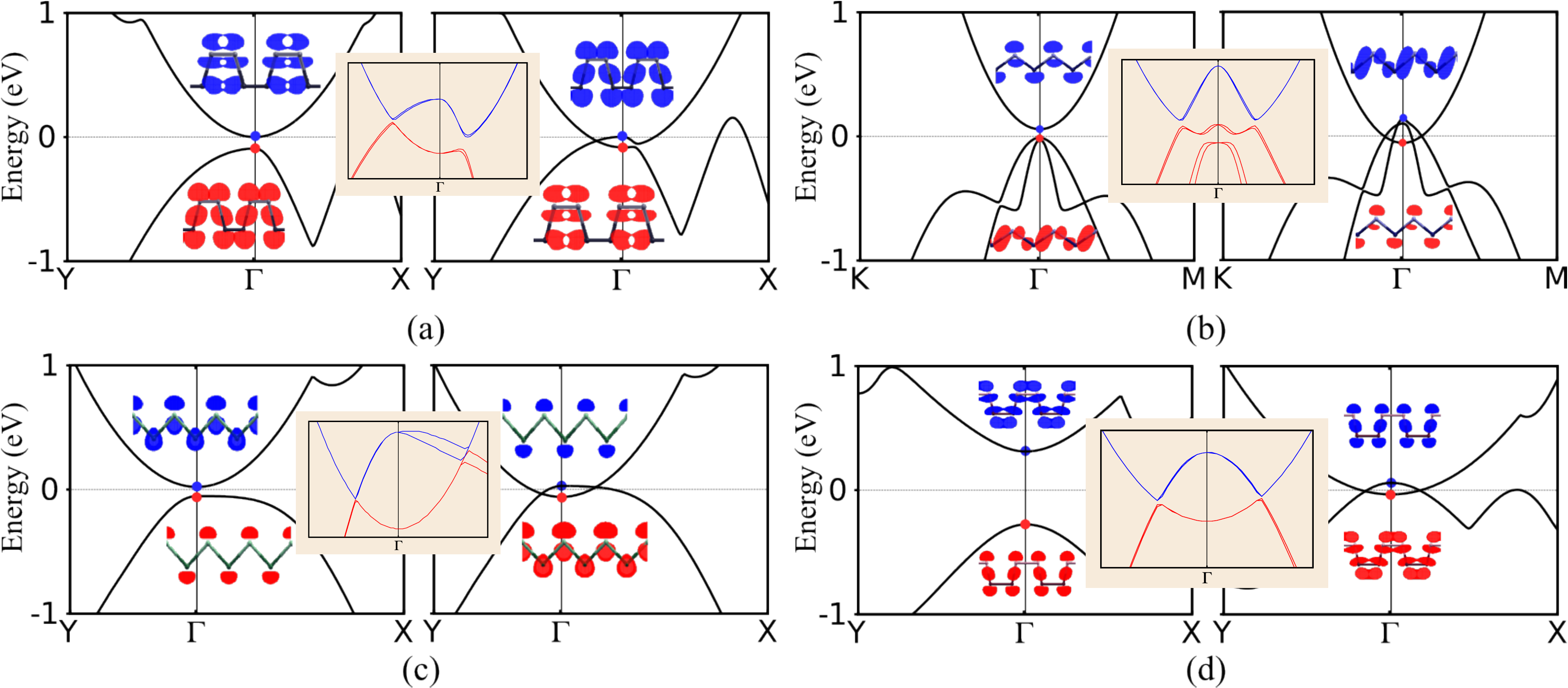}
\caption{Contribution of a particular wave-function (at the $\Gamma$ point, for band energies shown by red and blue dot) to the charge density for (a) $\alpha$-As (for $1.2$ V/\AA~and $1.5$ V/\AA) , (b) $\beta$-As (for $1.2$ V/\AA~and $1.5$ V/\AA) , (c) $\gamma$-As ($0.7$ V/\AA~ and $1.0$ V/\AA) and (d) $\delta$-As ($0.6$ V/\AA~and $0.9$ V/\AA). The left and right hand side panel in each figure represents the case just before and after the band inversion at finite electric field. This band inversion at the $\Gamma$-point, in all four allotropes ensures a nontrivial topological invariant $Z_2 = 1$ and thus, achieve the topological phase transition tuned by a transverse electric field. In particular $\alpha$-As becomes a topological metal, and all other phases become topological insulators beyond a critical electric field. The inset in all four panels, displays the spin resolved bands in the vicinity of the $\Gamma$ point, with spin-orbit coupling turned on, after the band inversion at finite electric filed . The opening of the bandgap, in all four insets is evident. This is a direct consequence of the broken inversion symmetry (on account of external field), in presence of spin orbit coupling. The values of the SOC gap along the $\Gamma Y$ direction is 6 meV, 1.8 meV, and 6 meV in panels (a), (c) and (d), respectively and 15 meV along the $\Gamma K$ direction in panel (b). 
\label{f4}}
\end{center}
\end{figure*}

Comparison between arsenic and it's predecessor phosphorous reveals several interesting facts. Due to their similarity of crystal structure and valence electron configuration, it is not surprising that we find some common features in electronic band structure of arsenic and phosphorous allotropes. For example, the energy dispersion of valence and conduction band are comparable for similar allotropes belonging to arsenic and phosphorous family~\cite{BP-Qiao,PRL.113.046804,zhu14}. Although the valence and conduction band energies are very close to the respective maximum and minimum at multiple points in the Brillouin zone for both the families, the VBM and CBM is located at different points for equivalent arsenic and phosphorous allotropes. As a result, we find that unlike phosphorous, $\alpha$ and $\delta$ allotrope of arsenic are indirect band gap semiconductors and $\gamma$ allotrope is direct bandgap semiconductor. Only the $\beta$-allotrope of arsenic and phosphorous~\cite{zhu14,Ghosh1} are qualitatively similar; both of them are indirect band gap semiconductors, although the location of VBM is different ($\Gamma$ point in arsenic and in the middle of $\Gamma$K direction in phosphorene).

We now discuss the impact of a transverse electric field and in-plane uniaxial strain on the bandstructure of all four monolayer As allotropes in the next two sections.  

\section{Impact of a transverse electric field}
\label{SecIV}

Band gap engineering in layered materials via the application of a transverse electric field has been demonstrated in bilayer graphene, silicene and phosphorene etc. Motivated by these studies, in this section we investigate the impact of a transverse electric field on the electronic properties of all four allotropes of monolayer arsenic.  In an experimental scenario, such an electric field is equivalent to the gate voltage applied in field effect transistors. The calculation is performed by doing the structural and electronic relaxation in presence of a sawtooth like potential applied in the $\hat{z}$ direction, perpendicular to the monolayer. The evolution of electronic band structure with changing electric field is shown in Fig.~\ref{f1}(i)-(l) and the evolution of the bandgap with increasing electric field strength is plotted in Fig.~\ref{f3}. We observe a transition from indirect bandgap (with larger $\Delta$) to direct bandgap (with smaller $\Delta$) semiconductor in case of $\beta$ and $\delta$-As beyond a threshold electric field of 0.9 and 0.6 V/\AA~, respectively. Although $\Delta$ decreases beyond certain electric field for $\alpha$ (0.8 V/\AA) and $\gamma$-As (0.5 V/\AA) also, but we do not observe any indirect to direct bandgap transition or vice versa [see Fig.~\ref{f3}]. 

{On increasing the electric field further, the bandgap vanishes for all of the four allotropes beyond a critical value; being maximum and minimum in $\beta$-As (1.2 V/\AA) and $\delta$-As (0.7 V/\AA), respectively. Interestingly, the energy gap between the original valence and conduction band (measured in zero field) does not decrease with increasing electric field. However, an unoccupied band (which is significantly higher in energy than the conduction band, when measured in zero field) decreases in energy with increasing electric field and ultimately comes down below the original conduction band beyond a critical field strength, eventually becoming the lowest unoccupied band. Thus, for all four As-allotropes the bandgap is found to be insensitive to the field strength till the crossover electric field is reached, and then decreases rapidly at higher field beyond the critical field strength [see Fig~\ref{f3}].} 
Note that, as the DFT-PBE method underestimates the bandgap, the actual strength of the electric field required for indirect to direct bandgap and semiconductor to metal transition is expected to be higher. However, we anticipate that as in the case of phosphorene and as shown for $\beta$-As in Ref.~[\onlinecite{Zhu_As}], multilayered structures of these allotropes will have a smaller bandgap, and will also undergo a similar metal insulator transition, albeit at smaller strengths of the electric fields. 

Interestingly, we find that in three of the monolayer allotropes: $\beta$-As, $\gamma$-As and $\delta$-As,  the semiconductor-metal transition for a transverse field strength of $1.3,~0.8 ~{\rm and}~ 0.7$ V/\AA ~respectively,  is accompanied by a band inversion between the valence and conduction band at the non-degenerate and time reversal invariant $\Gamma$ point. For $\alpha$-As monolayer, at external filed of $1.1$ V/\AA~the valance band crosses the Fermi energy at some point along the $\Gamma X$ direction, turning it into a normal metallic state.  However on further increase in the electric field beyond $1.3$ V/\AA~ there is a band inversion at the $\Gamma$-point in $\alpha$-As while it is in the metallic state. The band inversion in all four allotropes is confirmed by plotting the contribution of the wave-function, corresponding to the valence and conduction band at the $\Gamma$ point, to the total charge density. Comparing the wave-functions just before and after the transition, as shown in Fig.~\ref{f4}, the band inversion is clearly evident for all four allotropes. Moreover, after band inversion happening at finite electric field (which breaks inversion symmetry),  a gap opens up at the pair of points where the bands cross each other, when we switch on the spin orbit coupling [see inset of Fig.~\ref{f4}(a)-(d)]. Similar band inversion has also been reported in few layers of black phosphorene,\cite{Qihang} which also leads to topological phase transition. 

Motivated by the fact that band inversion is a precursor to topological phase transition, and that it occurs at the time reversal invariant $\Gamma$ point which is non-degenerate, we calculate the $Z_2$ topological invariant \cite{z2kane, z2Rahul2, z2Rahul, PhysRevB.83.235401} before and after the band inversion in all four allotropes. This confirms that  the $\beta$, $\gamma$ and $\delta$ allotropes of monolayer arsenic undergo a topological  phase transition from a normal insulator ($Z_2 =0$) to a topological insulator ($Z_2 =1$), with increasing transverse electric field, while $\alpha$-As transforms from a normal metal ($Z_2 =0$) to a topological metal ($Z_2 =1$). The $Z_2$ index was calculated using the $Z_2$-pack \cite{Z2Pack, PhysRevB.83.235401}.

Note that all the four monolayer allotropes have a relatively large bandgap, and thus require electric fields $> 0.7$ V/\AA ~(1.2 V/\AA ~for $\beta$-As), which are currently difficult to achieve experimentally. As a possible way of reducing the large critical electric field needed to explore the topological phase transition, multilayered allotropes of arsenic can be a suitable candidate since the bandgap is expected to decrease with increasing number of layers, similar to the case of multilayered black phosphorene \cite{phosphorene1}, and this has already been demonstrated for $\beta$-As layers \cite{Zhu_As}. Thus, the multilayered allotropes are expected to show metal insulator transition, accompanied by band inversion, and possibly a topological phase transition at much smaller critical electric fields on account of the reduced band gap. Such topoplogical phase transition has already been demonstrated for multilayered $\alpha$-P (or black phosphorene) in Ref.~[\onlinecite{Qihang}].

{An alternate way to emulate the transverse electric field in layered materials, is via the giant stark effect caused by potassium doping \cite{Kim723,Seung}. This has been recently used experimentally to drive trilayer black phosphorene into an anisotropic Dirac semi-metal state \cite{Kim723}. We anticipate that a similar experiment with controlled potassium doping in As-allotropes can be possibly used to probe the semiconductor-metal-topological phase transition.} 

\section{Strain-engineering of the Bandstructure}
\label{SecV}

\begin{figure}
\begin{center}
\includegraphics[width=1.0 \linewidth]{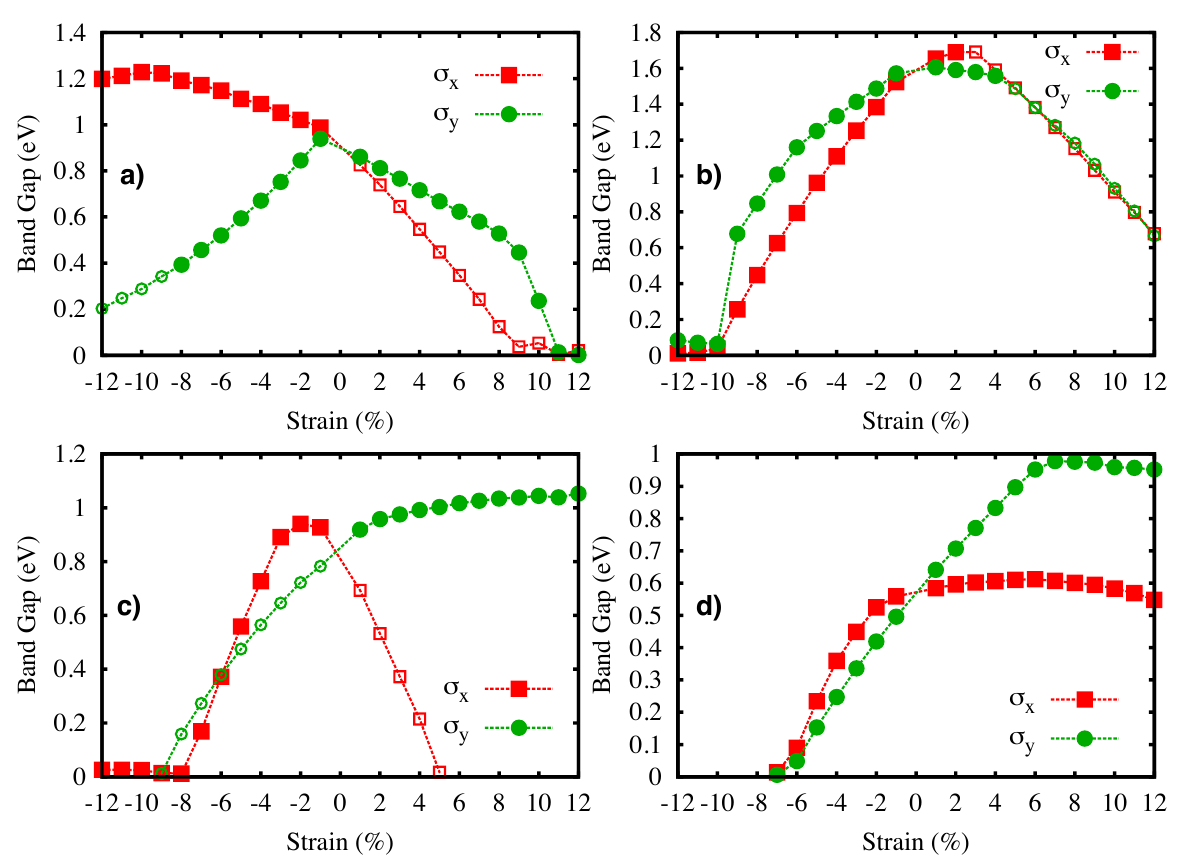}
\caption{Bandgap variation of monolayer arsenic allotropes is plotted as a function of uniaxial strain, separately applied in $x$ and $y$ direction; (a) $\alpha$-As, (b) $\beta$-As, (c) $\gamma$-As and (d) $\delta$-As. The open (filled) circles represent a direct (indirect) bandgap semiconductor. 
\label{f5}}
\end{center}
\end{figure}
Other than electric field, strain engineered band structure modification is also observed in various 2D materials; for example semi-metallic (zero bandgap) graphene can be converted to a semiconductor by applying strain.\cite{zhong} Electronic band structure of monolayer P allotropes, which are more closely related to arsenic allotropes, have a very complex dependence on applied strain, showing direct to indirect (or vice versa) bandgap semiconductor and semiconductor to metal transition.\cite{PRL.113.046804,zhu14} In case of arsenic, we study the effect of uniaxial compressive, as well as tensile strain, separately along the $x$ and $y$ direction, denoted by $\sigma_x$ and $\sigma_y$, respectively. Under the externally applied uniaxial strain, atomic positions are relaxed, followed by electronic band structure calculation for each of the allotropes and the variation of bandgap is plotted in Fig.~\ref{f5}. We observe that, although bandgap is very sensitive to the applied strain, but each of the allotropes remains a semiconductor over a large range of $\sigma$ values. 

Energetically more stable allotropes like $\alpha$ and $\beta$-arsenic undergo semiconductor to metal transition at very high value of applied strain ($\sim 10$\% tensile and compressive, respectively), while it takes slightly less amount of $\sigma$ ($\sim 5-6$\%) to induce a similar transition in $\gamma$ and $\delta$-arsenic. As a consistency check we note that for the case of $\beta$-As [see Fig.~\ref{f5}(b)], our results are qualitatively similar to those reported in Fig. 2(d) of Ref.~[\onlinecite{Zhu_As}]. {Note that, dependence of bandgap on strain shows a very complicated behavior and no general trend is observed, that fits all the allotropes. For example, in $\alpha$-As, bandgap increases for compressive $\sigma_x$, while it decreases for compressive $\sigma_y$ and tensile uniaxial strain in both directions. Similarly, for $\gamma$-As, bandgap changes hardly due to tensile $\sigma_y$, while it decreases for tensile $\sigma_x$, as well as compressive strain in both the directions. In case of $\delta$-As, while bandgap decreases due to compressive strain, it increases for tensile $\sigma_y$ but remains almost constant for tensile $\sigma_x$. Only for the case of $\beta$-As, which has the highest symmetry among the four allotropes, bandgap decreases for tensile, as well as compressive strain in both the directions. It should be noted that, strain changes the overlap of atomic orbitals, leading to bandgap modification. A more detail study on how the hopping integral changes as a function of strain, in conjunction with a tight binding model is required for a complete understanding of strain dependence of electronic bandstructure of 2D arsenic allotropes.} 
However, since the DFT-PBE method underestimates the  bandgap, the actual transition is expected to occur at even higher values of $\sigma$, which is possibly well beyond the limit of strain that can be applied in a material. Thus, for all practical purpose monolayer arsenic can be used as a semiconductor, with a possibility of bandgap tuning by strain.  Finally we note that while large values of uniaxial strain do induce metal-insulator transition in all four monolayer As allotropes, unlike the transverse electric field case this transition is not accompanied by a topological phase transition.

\section{Conclusion}
\label{SecVI}
To summarize, based on density functional calculations we predict four thermodynamically stable monolayer allotropes of As, similar to that of monolayer P allotropes \cite{PRL.113.046804}. We calculate the structural and electronic properties of all four allotropes, and our results for the bond length, monolayer height, cohesive energy, bandgap etc. for $\alpha$-As and $\beta$-As are in agreement with earlier works. Further we show that all four monolayer allotropes undergo a metal insulator transition on application of uniaxial in-plane strain, or a transverse electric field. More interestingly, the bandgap closure via transverse electric field is also accompanied by a band inversion around the $\Gamma$-point and a topological phase transition which supports spin separated gapless edge states. 
However on account of large bandgap in all four monolayer allotropes of As, the critical electric field required to drive the topological phase transition is large ($> 0.7$ V/\AA) and thus difficult to achieve experimentally. We anticipate that as in the case of multi-layered black phosphorene, multilayered allotropes of As may also undergo a topological phase transition, though at much smaller electric fields.  

\section*{Acknowledgements}  
AA acknowledges funding from the DST INSPIRE Faculty Award. SB acknowledges funding from SERB Fast Track Scheme for Young Scientist. We also thank CC IITK for providing HPC facility. Crystal structures illustrated in this paper are drawn using XCrySDen software.\cite{xc1}
\bibliography{Arsanene_ref}

\begin{thebibliography}{53}%
\makeatletter
\providecommand \@ifxundefined [1]{%
 \@ifx{#1\undefined}
}%
\providecommand \@ifnum [1]{%
 \ifnum #1\expandafter \@firstoftwo
 \else \expandafter \@secondoftwo
 \fi
}%
\providecommand \@ifx [1]{%
 \ifx #1\expandafter \@firstoftwo
 \else \expandafter \@secondoftwo
 \fi
}%
\providecommand \natexlab [1]{#1}%
\providecommand \enquote  [1]{``#1''}%
\providecommand \bibnamefont  [1]{#1}%
\providecommand \bibfnamefont [1]{#1}%
\providecommand \citenamefont [1]{#1}%
\providecommand \href@noop [0]{\@secondoftwo}%
\providecommand \href [0]{\begingroup \@sanitize@url \@href}%
\providecommand \@href[1]{\@@startlink{#1}\@@href}%
\providecommand \@@href[1]{\endgroup#1\@@endlink}%
\providecommand \@sanitize@url [0]{\catcode `\\12\catcode `\$12\catcode
  `\&12\catcode `\#12\catcode `\^12\catcode `\_12\catcode `\%12\relax}%
\providecommand \@@startlink[1]{}%
\providecommand \@@endlink[0]{}%
\providecommand \url  [0]{\begingroup\@sanitize@url \@url }%
\providecommand \@url [1]{\endgroup\@href {#1}{\urlprefix }}%
\providecommand \urlprefix  [0]{URL }%
\providecommand \Eprint [0]{\href }%
\providecommand \doibase [0]{http://dx.doi.org/}%
\providecommand \selectlanguage [0]{\@gobble}%
\providecommand \bibinfo  [0]{\@secondoftwo}%
\providecommand \bibfield  [0]{\@secondoftwo}%
\providecommand \translation [1]{[#1]}%
\providecommand \BibitemOpen [0]{}%
\providecommand \bibitemStop [0]{}%
\providecommand \bibitemNoStop [0]{.\EOS\space}%
\providecommand \EOS [0]{\spacefactor3000\relax}%
\providecommand \BibitemShut  [1]{\csname bibitem#1\endcsname}%
\let\auto@bib@innerbib\@empty
\bibitem [{\citenamefont {Geim}\ and\ \citenamefont
  {Novoselov}(2007)}]{graphene1}%
  \BibitemOpen
  \bibfield  {author} {\bibinfo {author} {\bibfnamefont {Andre~K}\ \bibnamefont
  {Geim}}\ and\ \bibinfo {author} {\bibfnamefont {Konstantin~S}\ \bibnamefont
  {Novoselov}},\ }\bibfield  {title} {\enquote {\bibinfo {title} {The rise of
  graphene},}\ }\href {\doibase 10.1038/nmat1849} {\bibfield  {journal}
  {\bibinfo  {journal} {Nat. Mater.}\ }\textbf {\bibinfo {volume} {6}},\
  \bibinfo {pages} {183--191} (\bibinfo {year} {2007})}\BibitemShut {NoStop}%
\bibitem [{\citenamefont {Castro~Neto}\ \emph {et~al.}(2009)\citenamefont
  {Castro~Neto}, \citenamefont {Guinea}, \citenamefont {Peres}, \citenamefont
  {Novoselov},\ and\ \citenamefont {Geim}}]{RevModPhys.81.109}%
  \BibitemOpen
  \bibfield  {author} {\bibinfo {author} {\bibfnamefont {A.~H.}\ \bibnamefont
  {Castro~Neto}}, \bibinfo {author} {\bibfnamefont {F.}~\bibnamefont {Guinea}},
  \bibinfo {author} {\bibfnamefont {N.~M.~R.}\ \bibnamefont {Peres}}, \bibinfo
  {author} {\bibfnamefont {K.~S.}\ \bibnamefont {Novoselov}}, \ and\ \bibinfo
  {author} {\bibfnamefont {A.~K.}\ \bibnamefont {Geim}},\ }\bibfield  {title}
  {\enquote {\bibinfo {title} {The electronic properties of graphene},}\ }\href
  {\doibase 10.1103/RevModPhys.81.109} {\bibfield  {journal} {\bibinfo
  {journal} {Rev. Mod. Phys.}\ }\textbf {\bibinfo {volume} {81}},\ \bibinfo
  {pages} {109--162} (\bibinfo {year} {2009})}\BibitemShut {NoStop}%
\bibitem [{\citenamefont {Katsnelson}(2012)}]{9781139031080}%
  \BibitemOpen
  \bibfield  {author} {\bibinfo {author} {\bibfnamefont {Mikhail~I.}\
  \bibnamefont {Katsnelson}},\ }\href
  {http://dx.doi.org/10.1017/CBO9781139031080} {\emph {\bibinfo {title}
  {Graphene}}}\ (\bibinfo  {publisher} {Cambridge University Press},\ \bibinfo
  {year} {2012})\ \bibinfo {note} {cambridge Books Online}\BibitemShut
  {NoStop}%
\bibitem [{\citenamefont {Takeda}\ and\ \citenamefont
  {Shiraishi}(1994)}]{PhysRevB.50.14916}%
  \BibitemOpen
  \bibfield  {author} {\bibinfo {author} {\bibfnamefont {Kyozaburo}\
  \bibnamefont {Takeda}}\ and\ \bibinfo {author} {\bibfnamefont {Kenji}\
  \bibnamefont {Shiraishi}},\ }\bibfield  {title} {\enquote {\bibinfo {title}
  {Theoretical possibility of stage corrugation in si and ge analogs of
  graphite},}\ }\href {\doibase 10.1103/PhysRevB.50.14916} {\bibfield
  {journal} {\bibinfo  {journal} {Phys. Rev. B}\ }\textbf {\bibinfo {volume}
  {50}},\ \bibinfo {pages} {14916--14922} (\bibinfo {year} {1994})}\BibitemShut
  {NoStop}%
\bibitem [{\citenamefont {Guzm\'an-Verri}\ and\ \citenamefont {Lew
  Yan~Voon}(2007)}]{PhysRevB.76.075131}%
  \BibitemOpen
  \bibfield  {author} {\bibinfo {author} {\bibfnamefont {Gian~G.}\ \bibnamefont
  {Guzm\'an-Verri}}\ and\ \bibinfo {author} {\bibfnamefont {L.~C.}\
  \bibnamefont {Lew Yan~Voon}},\ }\bibfield  {title} {\enquote {\bibinfo
  {title} {Electronic structure of silicon-based nanostructures},}\ }\href
  {\doibase 10.1103/PhysRevB.76.075131} {\bibfield  {journal} {\bibinfo
  {journal} {Phys. Rev. B}\ }\textbf {\bibinfo {volume} {76}},\ \bibinfo
  {pages} {075131} (\bibinfo {year} {2007})}\BibitemShut {NoStop}%
\bibitem [{\citenamefont {Lalmi}\ \emph {et~al.}(2010)\citenamefont {Lalmi},
  \citenamefont {Oughaddou}, \citenamefont {Enriquez}, \citenamefont {Kara},
  \citenamefont {Vizzini}, \citenamefont {Ealet},\ and\ \citenamefont
  {Aufray}}]{1.3524215}%
  \BibitemOpen
  \bibfield  {author} {\bibinfo {author} {\bibfnamefont {Boubekeur}\
  \bibnamefont {Lalmi}}, \bibinfo {author} {\bibfnamefont {Hamid}\ \bibnamefont
  {Oughaddou}}, \bibinfo {author} {\bibfnamefont {Hanna}\ \bibnamefont
  {Enriquez}}, \bibinfo {author} {\bibfnamefont {Abdelkader}\ \bibnamefont
  {Kara}}, \bibinfo {author} {\bibfnamefont {Sebastien}\ \bibnamefont
  {Vizzini}}, \bibinfo {author} {\bibfnamefont {Benidicte}\ \bibnamefont
  {Ealet}}, \ and\ \bibinfo {author} {\bibfnamefont {Bernard}\ \bibnamefont
  {Aufray}},\ }\bibfield  {title} {\enquote {\bibinfo {title} {Epitaxial growth
  of a silicene sheet},}\ }\href {\doibase 10.1063/1.3524215} {\bibfield
  {journal} {\bibinfo  {journal} {Appl. Phys. Lett.}\ }\textbf {\bibinfo
  {volume} {97}},\ \bibinfo {eid} {223109} (\bibinfo {year}
  {2010})}\BibitemShut {NoStop}%
\bibitem [{\citenamefont {Vogt}\ \emph {et~al.}(2012)\citenamefont {Vogt},
  \citenamefont {De~Padova}, \citenamefont {Quaresima}, \citenamefont {Avila},
  \citenamefont {Frantzeskakis}, \citenamefont {Asensio}, \citenamefont
  {Resta}, \citenamefont {Ealet},\ and\ \citenamefont {Le~Lay}}]{silicene2}%
  \BibitemOpen
  \bibfield  {author} {\bibinfo {author} {\bibfnamefont {Patrick}\ \bibnamefont
  {Vogt}}, \bibinfo {author} {\bibfnamefont {Paola}\ \bibnamefont {De~Padova}},
  \bibinfo {author} {\bibfnamefont {Claudio}\ \bibnamefont {Quaresima}},
  \bibinfo {author} {\bibfnamefont {Jose}\ \bibnamefont {Avila}}, \bibinfo
  {author} {\bibfnamefont {Emmanouil}\ \bibnamefont {Frantzeskakis}}, \bibinfo
  {author} {\bibfnamefont {Maria~Carmen}\ \bibnamefont {Asensio}}, \bibinfo
  {author} {\bibfnamefont {Andrea}\ \bibnamefont {Resta}}, \bibinfo {author}
  {\bibfnamefont {B\'en\'edicte}\ \bibnamefont {Ealet}}, \ and\ \bibinfo
  {author} {\bibfnamefont {Guy}\ \bibnamefont {Le~Lay}},\ }\bibfield  {title}
  {\enquote {\bibinfo {title} {Silicene: Compelling experimental evidence for
  graphene like two-dimensional silicon},}\ }\href {\doibase
  10.1103/PhysRevLett.108.155501} {\bibfield  {journal} {\bibinfo  {journal}
  {Phys. Rev. Lett.}\ }\textbf {\bibinfo {volume} {108}},\ \bibinfo {pages}
  {155501} (\bibinfo {year} {2012})}\BibitemShut {NoStop}%
\bibitem [{\citenamefont {Chen}\ \emph {et~al.}(2012)\citenamefont {Chen},
  \citenamefont {Liu}, \citenamefont {Feng}, \citenamefont {He}, \citenamefont
  {Cheng}, \citenamefont {Ding}, \citenamefont {Meng}, \citenamefont {Yao},\
  and\ \citenamefont {Wu}}]{PhysRevLett.109.056804}%
  \BibitemOpen
  \bibfield  {author} {\bibinfo {author} {\bibfnamefont {Lan}\ \bibnamefont
  {Chen}}, \bibinfo {author} {\bibfnamefont {Cheng-Cheng}\ \bibnamefont {Liu}},
  \bibinfo {author} {\bibfnamefont {Baojie}\ \bibnamefont {Feng}}, \bibinfo
  {author} {\bibfnamefont {Xiaoyue}\ \bibnamefont {He}}, \bibinfo {author}
  {\bibfnamefont {Peng}\ \bibnamefont {Cheng}}, \bibinfo {author}
  {\bibfnamefont {Zijing}\ \bibnamefont {Ding}}, \bibinfo {author}
  {\bibfnamefont {Sheng}\ \bibnamefont {Meng}}, \bibinfo {author}
  {\bibfnamefont {Yugui}\ \bibnamefont {Yao}}, \ and\ \bibinfo {author}
  {\bibfnamefont {Kehui}\ \bibnamefont {Wu}},\ }\bibfield  {title} {\enquote
  {\bibinfo {title} {Evidence for dirac fermions in a honeycomb lattice based
  on silicon},}\ }\href {\doibase 10.1103/PhysRevLett.109.056804} {\bibfield
  {journal} {\bibinfo  {journal} {Phys. Rev. Lett.}\ }\textbf {\bibinfo
  {volume} {109}},\ \bibinfo {pages} {056804} (\bibinfo {year}
  {2012})}\BibitemShut {NoStop}%
\bibitem [{\citenamefont {Bianco}\ \emph {et~al.}(2013)\citenamefont {Bianco},
  \citenamefont {Butler}, \citenamefont {Jiang}, \citenamefont {Restrepo},
  \citenamefont {Windl},\ and\ \citenamefont {Goldberger}}]{germanane1}%
  \BibitemOpen
  \bibfield  {author} {\bibinfo {author} {\bibfnamefont {Elisabeth}\
  \bibnamefont {Bianco}}, \bibinfo {author} {\bibfnamefont {Sheneve}\
  \bibnamefont {Butler}}, \bibinfo {author} {\bibfnamefont {Shishi}\
  \bibnamefont {Jiang}}, \bibinfo {author} {\bibfnamefont {Oscar~D.}\
  \bibnamefont {Restrepo}}, \bibinfo {author} {\bibfnamefont {Wolfgang}\
  \bibnamefont {Windl}}, \ and\ \bibinfo {author} {\bibfnamefont {Joshua~E.}\
  \bibnamefont {Goldberger}},\ }\bibfield  {title} {\enquote {\bibinfo {title}
  {Stability and exfoliation of germanane: A germanium graphane analogue},}\
  }\href {\doibase 10.1021/nn4009406} {\bibfield  {journal} {\bibinfo
  {journal} {ACS Nano}\ }\textbf {\bibinfo {volume} {7}},\ \bibinfo {pages}
  {4414--4421} (\bibinfo {year} {2013})}\BibitemShut {NoStop}%
\bibitem [{\citenamefont {Cahangirov}\ \emph {et~al.}(2009)\citenamefont
  {Cahangirov}, \citenamefont {Topsakal}, \citenamefont {Akt\"urk},
  \citenamefont {\ifmmode~\mbox{\c{S}}\else \c{S}\fi{}ahin},\ and\
  \citenamefont {Ciraci}}]{PhysRevLett.102.236804}%
  \BibitemOpen
  \bibfield  {author} {\bibinfo {author} {\bibfnamefont {S.}~\bibnamefont
  {Cahangirov}}, \bibinfo {author} {\bibfnamefont {M.}~\bibnamefont
  {Topsakal}}, \bibinfo {author} {\bibfnamefont {E.}~\bibnamefont {Akt\"urk}},
  \bibinfo {author} {\bibfnamefont {H.}~\bibnamefont
  {\ifmmode~\mbox{\c{S}}\else \c{S}\fi{}ahin}}, \ and\ \bibinfo {author}
  {\bibfnamefont {S.}~\bibnamefont {Ciraci}},\ }\bibfield  {title} {\enquote
  {\bibinfo {title} {Two- and one-dimensional honeycomb structures of silicon
  and germanium},}\ }\href {\doibase 10.1103/PhysRevLett.102.236804} {\bibfield
   {journal} {\bibinfo  {journal} {Phys. Rev. Lett.}\ }\textbf {\bibinfo
  {volume} {102}},\ \bibinfo {pages} {236804} (\bibinfo {year}
  {2009})}\BibitemShut {NoStop}%
\bibitem [{\citenamefont {Chhowalla}\ \emph {et~al.}(2013)\citenamefont
  {Chhowalla}, \citenamefont {Shin}, \citenamefont {Eda}, \citenamefont {Li},
  \citenamefont {Loh},\ and\ \citenamefont {~}}]{TMD1}%
  \BibitemOpen
  \bibfield  {author} {\bibinfo {author} {\bibfnamefont {Manish}\ \bibnamefont
  {Chhowalla}}, \bibinfo {author} {\bibfnamefont {Hyeon~Suk}\ \bibnamefont
  {Shin}}, \bibinfo {author} {\bibfnamefont {Goki}\ \bibnamefont {Eda}},
  \bibinfo {author} {\bibfnamefont {Lain-Jong}\ \bibnamefont {Li}}, \bibinfo
  {author} {\bibfnamefont {Kian~Ping}\ \bibnamefont {Loh}}, \ and\ \bibinfo
  {author} {\bibfnamefont {Hua}\ \bibnamefont {~}},\ }\bibfield  {title}
  {\enquote {\bibinfo {title} {The chemistry of two-dimensional layered
  transition metal dichalcogenide nanosheets},}\ }\href {\doibase
  10.1038/nchem.1589} {\bibfield  {journal} {\bibinfo  {journal} {Nat. Chem.}\
  }\textbf {\bibinfo {volume} {5}},\ \bibinfo {pages} {263--275} (\bibinfo
  {year} {2013})}\BibitemShut {NoStop}%
\bibitem [{\citenamefont {Xu}\ \emph {et~al.}(2014)\citenamefont {Xu},
  \citenamefont {Yao}, \citenamefont {Xiao},\ and\ \citenamefont
  {Heinz}}]{10.1038/nphys2942}%
  \BibitemOpen
  \bibfield  {author} {\bibinfo {author} {\bibfnamefont {Xiaodong}\
  \bibnamefont {Xu}}, \bibinfo {author} {\bibfnamefont {Wang}\ \bibnamefont
  {Yao}}, \bibinfo {author} {\bibfnamefont {Di}~\bibnamefont {Xiao}}, \ and\
  \bibinfo {author} {\bibfnamefont {Tony~F.}\ \bibnamefont {Heinz}},\
  }\bibfield  {title} {\enquote {\bibinfo {title} {Spin and pseudospins in
  layered transition metal dichalcogenides},}\ }\href
  {http://dx.doi.org/10.1038/nphys2942} {\bibfield  {journal} {\bibinfo
  {journal} {Nature Phys.}\ }\textbf {\bibinfo {volume} {10}},\ \bibinfo
  {pages} {343--350} (\bibinfo {year} {2014})}\BibitemShut {NoStop}%
\bibitem [{\citenamefont {Rastogi}\ \emph {et~al.}(2014)\citenamefont
  {Rastogi}, \citenamefont {Kumar}, \citenamefont {Bhowmick}, \citenamefont
  {Agarwal},\ and\ \citenamefont {Chauhan}}]{MoS2_priyank}%
  \BibitemOpen
  \bibfield  {author} {\bibinfo {author} {\bibfnamefont {Priyank}\ \bibnamefont
  {Rastogi}}, \bibinfo {author} {\bibfnamefont {Sanjay}\ \bibnamefont {Kumar}},
  \bibinfo {author} {\bibfnamefont {Somnath}\ \bibnamefont {Bhowmick}},
  \bibinfo {author} {\bibfnamefont {Amit}\ \bibnamefont {Agarwal}}, \ and\
  \bibinfo {author} {\bibfnamefont {Yogesh~Singh}\ \bibnamefont {Chauhan}},\
  }\bibfield  {title} {\enquote {\bibinfo {title} {Doping strategies for
  monolayer mos2 via surface adsorption: A systematic study},}\ }\href
  {\doibase 10.1021/jp510662n} {\bibfield  {journal} {\bibinfo  {journal} {The
  Journal of Physical Chemistry C}\ }\textbf {\bibinfo {volume} {118}},\
  \bibinfo {pages} {30309--30314} (\bibinfo {year} {2014})}\BibitemShut
  {NoStop}%
\bibitem [{\citenamefont {Li}\ \emph {et~al.}(2014)\citenamefont {Li},
  \citenamefont {Yu}, \citenamefont {Ye}, \citenamefont {Ge}, \citenamefont
  {Ou}, \citenamefont {Wu}, \citenamefont {Feng}, \citenamefont {Chen},\ and\
  \citenamefont {Zhang}}]{BP-transistor}%
  \BibitemOpen
  \bibfield  {author} {\bibinfo {author} {\bibfnamefont {Likai}\ \bibnamefont
  {Li}}, \bibinfo {author} {\bibfnamefont {Yijun}\ \bibnamefont {Yu}}, \bibinfo
  {author} {\bibfnamefont {Guo~Jun}\ \bibnamefont {Ye}}, \bibinfo {author}
  {\bibfnamefont {Qingqin}\ \bibnamefont {Ge}}, \bibinfo {author}
  {\bibfnamefont {Xuedong}\ \bibnamefont {Ou}}, \bibinfo {author}
  {\bibfnamefont {Hua}\ \bibnamefont {Wu}}, \bibinfo {author} {\bibfnamefont
  {Donglai}\ \bibnamefont {Feng}}, \bibinfo {author} {\bibfnamefont {Xian~Hui}\
  \bibnamefont {Chen}}, \ and\ \bibinfo {author} {\bibfnamefont {Yuanbo}\
  \bibnamefont {Zhang}},\ }\bibfield  {title} {\enquote {\bibinfo {title}
  {Black phosphorus field-effect transistors},}\ }\href {\doibase
  10.1038/nnano.2014.35} {\bibfield  {journal} {\bibinfo  {journal} {Nat.
  Nanotechnol.}\ }\textbf {\bibinfo {volume} {9}},\ \bibinfo {pages} {372--377}
  (\bibinfo {year} {2014})}\BibitemShut {NoStop}%
\bibitem [{\citenamefont {Liu}\ \emph {et~al.}(2014)\citenamefont {Liu},
  \citenamefont {Neal}, \citenamefont {Zhu}, \citenamefont {Luo}, \citenamefont
  {Xu}, \citenamefont {Tom\'anek},\ and\ \citenamefont {Ye}}]{phosphorene1}%
  \BibitemOpen
  \bibfield  {author} {\bibinfo {author} {\bibfnamefont {Han}\ \bibnamefont
  {Liu}}, \bibinfo {author} {\bibfnamefont {Adam~T.}\ \bibnamefont {Neal}},
  \bibinfo {author} {\bibfnamefont {Zhen}\ \bibnamefont {Zhu}}, \bibinfo
  {author} {\bibfnamefont {Zhe}\ \bibnamefont {Luo}}, \bibinfo {author}
  {\bibfnamefont {Xianfan}\ \bibnamefont {Xu}}, \bibinfo {author}
  {\bibfnamefont {David}\ \bibnamefont {Tom\'anek}}, \ and\ \bibinfo {author}
  {\bibfnamefont {Peide~D.}\ \bibnamefont {Ye}},\ }\bibfield  {title} {\enquote
  {\bibinfo {title} {Phosphorene: An unexplored 2d semiconductor with a high
  hole mobility},}\ }\href {\doibase 10.1021/nn501226z} {\bibfield  {journal}
  {\bibinfo  {journal} {ACS Nano}\ }\textbf {\bibinfo {volume} {8}},\ \bibinfo
  {pages} {4033--4041} (\bibinfo {year} {2014})}\BibitemShut {NoStop}%
\bibitem [{\citenamefont {Koenig}\ \emph {et~al.}(2014)\citenamefont {Koenig},
  \citenamefont {Doganov}, \citenamefont {Schmidt}, \citenamefont
  {Castro~Neto},\ and\ \citenamefont {Ozyilmaz}}]{APL_Neto}%
  \BibitemOpen
  \bibfield  {author} {\bibinfo {author} {\bibfnamefont {Steven~P.}\
  \bibnamefont {Koenig}}, \bibinfo {author} {\bibfnamefont {Rostislav~A.}\
  \bibnamefont {Doganov}}, \bibinfo {author} {\bibfnamefont {Hennrik}\
  \bibnamefont {Schmidt}}, \bibinfo {author} {\bibfnamefont {A.~H.}\
  \bibnamefont {Castro~Neto}}, \ and\ \bibinfo {author} {\bibfnamefont
  {Barbaros}\ \bibnamefont {Ozyilmaz}},\ }\bibfield  {title} {\enquote
  {\bibinfo {title} {Electric field effect in ultrathin black phosphorus},}\
  }\href {\doibase 10.1063/1.4868132} {\bibfield  {journal} {\bibinfo
  {journal} {Appl. Phys. Lett.}\ }\textbf {\bibinfo {volume} {104}},\ \bibinfo
  {eid} {103106} (\bibinfo {year} {2014})}\BibitemShut {NoStop}%
\bibitem [{\citenamefont {Schwierz}(2010)}]{graphene_tr}%
  \BibitemOpen
  \bibfield  {author} {\bibinfo {author} {\bibfnamefont {Frank}\ \bibnamefont
  {Schwierz}},\ }\bibfield  {title} {\enquote {\bibinfo {title} {Graphene
  transistors},}\ }\href {\doibase 10.1038/nnano.2010.89} {\bibfield  {journal}
  {\bibinfo  {journal} {Nat. Nanotechnol.}\ }\textbf {\bibinfo {volume} {5}},\
  \bibinfo {pages} {487--496} (\bibinfo {year} {2010})}\BibitemShut {NoStop}%
\bibitem [{\citenamefont {Radisavljevic}\ \emph {et~al.}(2011)\citenamefont
  {Radisavljevic}, \citenamefont {A}, \citenamefont {J}, \citenamefont {V},\
  and\ \citenamefont {A}}]{MoS2_tr}%
  \BibitemOpen
  \bibfield  {author} {\bibinfo {author} {\bibfnamefont {B}~\bibnamefont
  {Radisavljevic}}, \bibinfo {author} {\bibfnamefont {Radenovic}\ \bibnamefont
  {A}}, \bibinfo {author} {\bibfnamefont {Brivio}\ \bibnamefont {J}}, \bibinfo
  {author} {\bibfnamefont {Giacometti}\ \bibnamefont {V}}, \ and\ \bibinfo
  {author} {\bibfnamefont {Kis}\ \bibnamefont {A}},\ }\bibfield  {title}
  {\enquote {\bibinfo {title} {Single-layer mos2 transistors},}\ }\href
  {\doibase 10.1038/nnano.2010.279} {\bibfield  {journal} {\bibinfo  {journal}
  {Nat. Nanotechnol.}\ }\textbf {\bibinfo {volume} {6}},\ \bibinfo {pages}
  {147--150} (\bibinfo {year} {2011})}\BibitemShut {NoStop}%
\bibitem [{\citenamefont {Qiao}\ \emph {et~al.}(2014)\citenamefont {Qiao},
  \citenamefont {Kong}, \citenamefont {Hu}, \citenamefont {Yang},\ and\
  \citenamefont {Ji}}]{BP-Qiao}%
  \BibitemOpen
  \bibfield  {author} {\bibinfo {author} {\bibfnamefont {Jingsi}\ \bibnamefont
  {Qiao}}, \bibinfo {author} {\bibfnamefont {Xianghua}\ \bibnamefont {Kong}},
  \bibinfo {author} {\bibfnamefont {Zhi-Xin}\ \bibnamefont {Hu}}, \bibinfo
  {author} {\bibfnamefont {Feng}\ \bibnamefont {Yang}}, \ and\ \bibinfo
  {author} {\bibfnamefont {Wei}\ \bibnamefont {Ji}},\ }\bibfield  {title}
  {\enquote {\bibinfo {title} {High-mobility transport anisotropy and linear
  dichroism in few-layer black phosphorus},}\ }\href
  {http://dx.doi.org/10.1038/ncomms5475} {\bibfield  {journal} {\bibinfo
  {journal} {Nat Commun}\ }\textbf {\bibinfo {volume} {5}} (\bibinfo {year}
  {2014})}\BibitemShut {NoStop}%
\bibitem [{\citenamefont {Rodin}\ \emph {et~al.}(2014)\citenamefont {Rodin},
  \citenamefont {Carvalho},\ and\ \citenamefont {Castro~Neto}}]{PRLNeto}%
  \BibitemOpen
  \bibfield  {author} {\bibinfo {author} {\bibfnamefont {A.~S.}\ \bibnamefont
  {Rodin}}, \bibinfo {author} {\bibfnamefont {A.}~\bibnamefont {Carvalho}}, \
  and\ \bibinfo {author} {\bibfnamefont {A.~H.}\ \bibnamefont {Castro~Neto}},\
  }\bibfield  {title} {\enquote {\bibinfo {title} {Strain-induced gap
  modification in black phosphorus},}\ }\href {\doibase
  10.1103/PhysRevLett.112.176801} {\bibfield  {journal} {\bibinfo  {journal}
  {Phys. Rev. Lett.}\ }\textbf {\bibinfo {volume} {112}},\ \bibinfo {pages}
  {176801} (\bibinfo {year} {2014})}\BibitemShut {NoStop}%
\bibitem [{\citenamefont {Zhu}\ and\ \citenamefont {Tom\'anek}(2014)}]{zhu14}%
  \BibitemOpen
  \bibfield  {author} {\bibinfo {author} {\bibfnamefont {Zhen}\ \bibnamefont
  {Zhu}}\ and\ \bibinfo {author} {\bibfnamefont {David}\ \bibnamefont
  {Tom\'anek}},\ }\bibfield  {title} {\enquote {\bibinfo {title}
  {Semiconducting layered blue phosphorus: A computational study},}\ }\href
  {\doibase 10.1103/PhysRevLett.112.176802} {\bibfield  {journal} {\bibinfo
  {journal} {Phys. Rev. Lett.}\ }\textbf {\bibinfo {volume} {112}},\ \bibinfo
  {pages} {176802} (\bibinfo {year} {2014})}\BibitemShut {NoStop}%
\bibitem [{\citenamefont {Guan}\ \emph {et~al.}(2014)\citenamefont {Guan},
  \citenamefont {Zhu},\ and\ \citenamefont {Tom\'anek}}]{PRL.113.046804}%
  \BibitemOpen
  \bibfield  {author} {\bibinfo {author} {\bibfnamefont {Jie}\ \bibnamefont
  {Guan}}, \bibinfo {author} {\bibfnamefont {Zhen}\ \bibnamefont {Zhu}}, \ and\
  \bibinfo {author} {\bibfnamefont {David}\ \bibnamefont {Tom\'anek}},\
  }\bibfield  {title} {\enquote {\bibinfo {title} {Phase coexistence and
  metal-insulator transition in few-layer phosphorene: A computational
  study},}\ }\href {\doibase 10.1103/PhysRevLett.113.046804} {\bibfield
  {journal} {\bibinfo  {journal} {Phys. Rev. Lett.}\ }\textbf {\bibinfo
  {volume} {113}},\ \bibinfo {pages} {046804} (\bibinfo {year}
  {2014})}\BibitemShut {NoStop}%
\bibitem [{\citenamefont {Ezawa}(2014)}]{BP-Ezawa}%
  \BibitemOpen
  \bibfield  {author} {\bibinfo {author} {\bibfnamefont {Motohiko}\
  \bibnamefont {Ezawa}},\ }\bibfield  {title} {\enquote {\bibinfo {title}
  {Topological origin of quasi-flat edge band in phosphorene},}\ }\href
  {http://stacks.iop.org/1367-2630/16/i=11/a=115004} {\bibfield  {journal}
  {\bibinfo  {journal} {New Journal of Physics}\ }\textbf {\bibinfo {volume}
  {16}},\ \bibinfo {pages} {115004} (\bibinfo {year} {2014})}\BibitemShut
  {NoStop}%
\bibitem [{\citenamefont {Ghosh}\ \emph {et~al.}(2015)\citenamefont {Ghosh},
  \citenamefont {Nahas}, \citenamefont {Bhowmick},\ and\ \citenamefont
  {Agarwal}}]{Ghosh1}%
  \BibitemOpen
  \bibfield  {author} {\bibinfo {author} {\bibfnamefont {Barun}\ \bibnamefont
  {Ghosh}}, \bibinfo {author} {\bibfnamefont {Suhas}\ \bibnamefont {Nahas}},
  \bibinfo {author} {\bibfnamefont {Somnath}\ \bibnamefont {Bhowmick}}, \ and\
  \bibinfo {author} {\bibfnamefont {Amit}\ \bibnamefont {Agarwal}},\ }\bibfield
   {title} {\enquote {\bibinfo {title} {Electric field induced gap modification
  in ultrathin blue phosphorus},}\ }\href {\doibase 10.1103/PhysRevB.91.115433}
  {\bibfield  {journal} {\bibinfo  {journal} {Phys. Rev. B}\ }\textbf {\bibinfo
  {volume} {91}},\ \bibinfo {pages} {115433} (\bibinfo {year}
  {2015})}\BibitemShut {NoStop}%
\bibitem [{\citenamefont {{Rastogi}}\ \emph {et~al.}(2015)\citenamefont
  {{Rastogi}}, \citenamefont {{Kumar}}, \citenamefont {{Bhowmick}},
  \citenamefont {{Agarwal}},\ and\ \citenamefont {{Singh
  Chauhan}}}]{Priyank_Padatom}%
  \BibitemOpen
  \bibfield  {author} {\bibinfo {author} {\bibfnamefont {P.}~\bibnamefont
  {{Rastogi}}}, \bibinfo {author} {\bibfnamefont {S.}~\bibnamefont {{Kumar}}},
  \bibinfo {author} {\bibfnamefont {S.}~\bibnamefont {{Bhowmick}}}, \bibinfo
  {author} {\bibfnamefont {A.}~\bibnamefont {{Agarwal}}}, \ and\ \bibinfo
  {author} {\bibfnamefont {Y.}~\bibnamefont {{Singh Chauhan}}},\ }\bibfield
  {title} {\enquote {\bibinfo {title} {{Effective Doping of Monolayer
  Phosphorene by Surface Adsorption of Atoms for Electronic and Spintronic
  Applications}},}\ }\href@noop {} {\bibfield  {journal} {\bibinfo  {journal}
  {ArXiv e-prints}\ } (\bibinfo {year} {2015})},\ \Eprint
  {http://arxiv.org/abs/1503.04296} {arXiv:1503.04296 [cond-mat.mes-hall]}
  \BibitemShut {NoStop}%
\bibitem [{\citenamefont {Nahas}\ \emph {et~al.}(2016)\citenamefont {Nahas},
  \citenamefont {Ghosh}, \citenamefont {Bhowmick},\ and\ \citenamefont
  {Agarwal}}]{Suhas_gs}%
  \BibitemOpen
  \bibfield  {author} {\bibinfo {author} {\bibfnamefont {Suhas}\ \bibnamefont
  {Nahas}}, \bibinfo {author} {\bibfnamefont {Barun}\ \bibnamefont {Ghosh}},
  \bibinfo {author} {\bibfnamefont {Somnath}\ \bibnamefont {Bhowmick}}, \ and\
  \bibinfo {author} {\bibfnamefont {Amit}\ \bibnamefont {Agarwal}},\ }\bibfield
   {title} {\enquote {\bibinfo {title} {First-principles cluster expansion
  study of functionalization of black phosphorene via fluorination and
  oxidation},}\ }\href {\doibase 10.1103/PhysRevB.93.165413} {\bibfield
  {journal} {\bibinfo  {journal} {Phys. Rev. B}\ }\textbf {\bibinfo {volume}
  {93}},\ \bibinfo {pages} {165413} (\bibinfo {year} {2016})}\BibitemShut
  {NoStop}%
\bibitem [{\citenamefont {Zhu}\ \emph {et~al.}(2015{\natexlab{a}})\citenamefont
  {Zhu}, \citenamefont {Guan},\ and\ \citenamefont {Tom\'anek}}]{Zhu_As}%
  \BibitemOpen
  \bibfield  {author} {\bibinfo {author} {\bibfnamefont {Zhen}\ \bibnamefont
  {Zhu}}, \bibinfo {author} {\bibfnamefont {Jie}\ \bibnamefont {Guan}}, \ and\
  \bibinfo {author} {\bibfnamefont {David}\ \bibnamefont {Tom\'anek}},\
  }\bibfield  {title} {\enquote {\bibinfo {title} {Strain-induced
  metal-semiconductor transition in monolayers and bilayers of gray arsenic: A
  computational study},}\ }\href {\doibase 10.1103/PhysRevB.91.161404}
  {\bibfield  {journal} {\bibinfo  {journal} {Phys. Rev. B}\ }\textbf {\bibinfo
  {volume} {91}},\ \bibinfo {pages} {161404} (\bibinfo {year}
  {2015}{\natexlab{a}})}\BibitemShut {NoStop}%
\bibitem [{\citenamefont {Kamal}\ and\ \citenamefont {Ezawa}(2015)}]{Kamal}%
  \BibitemOpen
  \bibfield  {author} {\bibinfo {author} {\bibfnamefont {C.}~\bibnamefont
  {Kamal}}\ and\ \bibinfo {author} {\bibfnamefont {Motohiko}\ \bibnamefont
  {Ezawa}},\ }\bibfield  {title} {\enquote {\bibinfo {title} {Arsenene:
  Two-dimensional buckled and puckered honeycomb arsenic systems},}\ }\href
  {\doibase 10.1103/PhysRevB.91.085423} {\bibfield  {journal} {\bibinfo
  {journal} {Phys. Rev. B}\ }\textbf {\bibinfo {volume} {91}},\ \bibinfo
  {pages} {085423} (\bibinfo {year} {2015})}\BibitemShut {NoStop}%
\bibitem [{\citenamefont {Zeraati}\ \emph {et~al.}(2016)\citenamefont
  {Zeraati}, \citenamefont {Vaez~Allaei}, \citenamefont
  {Abdolhosseini~Sarsari}, \citenamefont {Pourfath},\ and\ \citenamefont
  {Donadio}}]{Zerati}%
  \BibitemOpen
  \bibfield  {author} {\bibinfo {author} {\bibfnamefont {Majid}\ \bibnamefont
  {Zeraati}}, \bibinfo {author} {\bibfnamefont {S.~Mehdi}\ \bibnamefont
  {Vaez~Allaei}}, \bibinfo {author} {\bibfnamefont {I.}~\bibnamefont
  {Abdolhosseini~Sarsari}}, \bibinfo {author} {\bibfnamefont {Mahdi}\
  \bibnamefont {Pourfath}}, \ and\ \bibinfo {author} {\bibfnamefont {Davide}\
  \bibnamefont {Donadio}},\ }\bibfield  {title} {\enquote {\bibinfo {title}
  {Highly anisotropic thermal conductivity of arsenene: An \textit{ab initio}
  study},}\ }\href {\doibase 10.1103/PhysRevB.93.085424} {\bibfield  {journal}
  {\bibinfo  {journal} {Phys. Rev. B}\ }\textbf {\bibinfo {volume} {93}},\
  \bibinfo {pages} {085424} (\bibinfo {year} {2016})}\BibitemShut {NoStop}%
\bibitem [{\citenamefont {Cao}\ \emph {et~al.}(2015)\citenamefont {Cao},
  \citenamefont {Yu},\ and\ \citenamefont {Lu}}]{CaoSM}%
  \BibitemOpen
  \bibfield  {author} {\bibinfo {author} {\bibfnamefont {Huawei}\ \bibnamefont
  {Cao}}, \bibinfo {author} {\bibfnamefont {Zhongyuan}\ \bibnamefont {Yu}}, \
  and\ \bibinfo {author} {\bibfnamefont {Pengfei}\ \bibnamefont {Lu}},\
  }\bibfield  {title} {\enquote {\bibinfo {title} {Electronic properties of
  monolayer and bilayer arsenene under in-plain biaxial strains},}\ }\href
  {\doibase 10.1016/j.spmi.2015.08.006} {\bibfield  {journal} {\bibinfo
  {journal} {Superlattices and Microstructures}\ }\textbf {\bibinfo {volume}
  {86}},\ \bibinfo {pages} {501 -- 507} (\bibinfo {year} {2015})}\BibitemShut
  {NoStop}%
\bibitem [{\citenamefont {Zhang}\ \emph
  {et~al.}(2015{\natexlab{a}})\citenamefont {Zhang}, \citenamefont {Xie},
  \citenamefont {Yang}, \citenamefont {Wang}, \citenamefont {Si},\ and\
  \citenamefont {Xue}}]{Zhiya}%
  \BibitemOpen
  \bibfield  {author} {\bibinfo {author} {\bibfnamefont {Zhiya}\ \bibnamefont
  {Zhang}}, \bibinfo {author} {\bibfnamefont {Jiafeng}\ \bibnamefont {Xie}},
  \bibinfo {author} {\bibfnamefont {Dezheng}\ \bibnamefont {Yang}}, \bibinfo
  {author} {\bibfnamefont {Yuhua}\ \bibnamefont {Wang}}, \bibinfo {author}
  {\bibfnamefont {Mingsu}\ \bibnamefont {Si}}, \ and\ \bibinfo {author}
  {\bibfnamefont {Desheng}\ \bibnamefont {Xue}},\ }\bibfield  {title} {\enquote
  {\bibinfo {title} {Manifestation of unexpected semiconducting properties in
  few-layer orthorhombic arsenene},}\ }\href
  {http://stacks.iop.org/1882-0786/8/i=5/a=055201} {\bibfield  {journal}
  {\bibinfo  {journal} {Applied Physics Express}\ }\textbf {\bibinfo {volume}
  {8}},\ \bibinfo {pages} {055201} (\bibinfo {year}
  {2015}{\natexlab{a}})}\BibitemShut {NoStop}%
\bibitem [{\citenamefont {Zhang}\ \emph
  {et~al.}(2015{\natexlab{b}})\citenamefont {Zhang}, \citenamefont {Hu},
  \citenamefont {Hu}, \citenamefont {Cai},\ and\ \citenamefont
  {Zeng}}]{Shengli_APL}%
  \BibitemOpen
  \bibfield  {author} {\bibinfo {author} {\bibfnamefont {Shengli}\ \bibnamefont
  {Zhang}}, \bibinfo {author} {\bibfnamefont {Yonghong}\ \bibnamefont {Hu}},
  \bibinfo {author} {\bibfnamefont {Ziyu}\ \bibnamefont {Hu}}, \bibinfo
  {author} {\bibfnamefont {Bo}~\bibnamefont {Cai}}, \ and\ \bibinfo {author}
  {\bibfnamefont {Haibo}\ \bibnamefont {Zeng}},\ }\bibfield  {title} {\enquote
  {\bibinfo {title} {Hydrogenated arsenenes as planar magnet and dirac
  material},}\ }\href {\doibase 10.1063/1.4926761} {\bibfield  {journal}
  {\bibinfo  {journal} {Applied Physics Letters}\ }\textbf {\bibinfo {volume}
  {107}},\ \bibinfo {pages} {022102} (\bibinfo {year}
  {2015}{\natexlab{b}})}\BibitemShut {NoStop}%
\bibitem [{\citenamefont {ping Wang}\ \emph {et~al.}(2015)\citenamefont {ping
  Wang}, \citenamefont {wen Zhang}, \citenamefont {xiao Ji},\ and\
  \citenamefont {ji~Wang}}]{Yaping_APE}%
  \BibitemOpen
  \bibfield  {author} {\bibinfo {author} {\bibfnamefont {Ya}~\bibnamefont {ping
  Wang}}, \bibinfo {author} {\bibfnamefont {Chang}\ \bibnamefont {wen Zhang}},
  \bibinfo {author} {\bibfnamefont {Wei}\ \bibnamefont {xiao Ji}}, \ and\
  \bibinfo {author} {\bibfnamefont {Pei}\ \bibnamefont {ji~Wang}},\ }\bibfield
  {title} {\enquote {\bibinfo {title} {Unexpected band structure and half-metal
  in non-metal-doped arsenene sheet},}\ }\href
  {http://stacks.iop.org/1882-0786/8/i=6/a=065202} {\bibfield  {journal}
  {\bibinfo  {journal} {Applied Physics Express}\ }\textbf {\bibinfo {volume}
  {8}},\ \bibinfo {pages} {065202} (\bibinfo {year} {2015})}\BibitemShut
  {NoStop}%
\bibitem [{\citenamefont {Zhu}\ \emph {et~al.}(2015{\natexlab{b}})\citenamefont
  {Zhu}, \citenamefont {Guan},\ and\ \citenamefont {Tom{\'a}nek}}]{ZhenZhu_NL}%
  \BibitemOpen
  \bibfield  {author} {\bibinfo {author} {\bibfnamefont {Zhen}\ \bibnamefont
  {Zhu}}, \bibinfo {author} {\bibfnamefont {Jie}\ \bibnamefont {Guan}}, \ and\
  \bibinfo {author} {\bibfnamefont {David}\ \bibnamefont {Tom{\'a}nek}},\
  }\bibfield  {title} {\enquote {\bibinfo {title} {Structural transition in
  layered as1--xpx compounds: A computational study},}\ }\href {\doibase
  10.1021/acs.nanolett.5b02227} {\bibfield  {journal} {\bibinfo  {journal}
  {Nano Letters}\ }\textbf {\bibinfo {volume} {15}},\ \bibinfo {pages}
  {6042--6046} (\bibinfo {year} {2015}{\natexlab{b}})}\BibitemShut {NoStop}%
\bibitem [{\citenamefont {Zhang}\ \emph
  {et~al.}(2015{\natexlab{c}})\citenamefont {Zhang}, \citenamefont {Cao},
  \citenamefont {Zhang}, \citenamefont {Wang}, \citenamefont {Xue},\ and\
  \citenamefont {Si}}]{Zhang_AIPA}%
  \BibitemOpen
  \bibfield  {author} {\bibinfo {author} {\bibfnamefont {Z.~Y.}\ \bibnamefont
  {Zhang}}, \bibinfo {author} {\bibfnamefont {H.~N.}\ \bibnamefont {Cao}},
  \bibinfo {author} {\bibfnamefont {J.~C.}\ \bibnamefont {Zhang}}, \bibinfo
  {author} {\bibfnamefont {Y.~H.}\ \bibnamefont {Wang}}, \bibinfo {author}
  {\bibfnamefont {D.~S.}\ \bibnamefont {Xue}}, \ and\ \bibinfo {author}
  {\bibfnamefont {M.~S.}\ \bibnamefont {Si}},\ }\bibfield  {title} {\enquote
  {\bibinfo {title} {Orientation and strain modulated electronic structures in
  puckered arsenene nanoribbons},}\ }\href {\doibase 10.1063/1.4922329}
  {\bibfield  {journal} {\bibinfo  {journal} {AIP Advances}\ }\textbf {\bibinfo
  {volume} {5}},\ \bibinfo {pages} {067117} (\bibinfo {year}
  {2015}{\natexlab{c}})}\BibitemShut {NoStop}%
\bibitem [{\citenamefont {Han}\ \emph {et~al.}(2015)\citenamefont {Han},
  \citenamefont {Xie}, \citenamefont {Zhang}, \citenamefont {Yang},
  \citenamefont {Si},\ and\ \citenamefont {Xue}}]{Han_APE}%
  \BibitemOpen
  \bibfield  {author} {\bibinfo {author} {\bibfnamefont {Jianwei}\ \bibnamefont
  {Han}}, \bibinfo {author} {\bibfnamefont {Jiafeng}\ \bibnamefont {Xie}},
  \bibinfo {author} {\bibfnamefont {Zhiya}\ \bibnamefont {Zhang}}, \bibinfo
  {author} {\bibfnamefont {Dezheng}\ \bibnamefont {Yang}}, \bibinfo {author}
  {\bibfnamefont {Mingsu}\ \bibnamefont {Si}}, \ and\ \bibinfo {author}
  {\bibfnamefont {Desheng}\ \bibnamefont {Xue}},\ }\bibfield  {title} {\enquote
  {\bibinfo {title} {Negative poisson's ratios in few-layer orthorhombic
  arsenic: First-principles calculations},}\ }\href
  {http://stacks.iop.org/1882-0786/8/i=4/a=041801} {\bibfield  {journal}
  {\bibinfo  {journal} {Applied Physics Express}\ }\textbf {\bibinfo {volume}
  {8}},\ \bibinfo {pages} {041801} (\bibinfo {year} {2015})}\BibitemShut
  {NoStop}%
\bibitem [{\citenamefont {Kou}\ \emph {et~al.}(2015)\citenamefont {Kou},
  \citenamefont {Ma}, \citenamefont {Tan}, \citenamefont {Frauenheim},
  \citenamefont {Du},\ and\ \citenamefont {Smith}}]{Kou_JPCC}%
  \BibitemOpen
  \bibfield  {author} {\bibinfo {author} {\bibfnamefont {Liangzhi}\
  \bibnamefont {Kou}}, \bibinfo {author} {\bibfnamefont {Yandong}\ \bibnamefont
  {Ma}}, \bibinfo {author} {\bibfnamefont {Xin}\ \bibnamefont {Tan}}, \bibinfo
  {author} {\bibfnamefont {Thomas}\ \bibnamefont {Frauenheim}}, \bibinfo
  {author} {\bibfnamefont {Aijun}\ \bibnamefont {Du}}, \ and\ \bibinfo {author}
  {\bibfnamefont {Sean}\ \bibnamefont {Smith}},\ }\bibfield  {title} {\enquote
  {\bibinfo {title} {Structural and electronic properties of layered arsenic
  and antimony arsenide},}\ }\href {\doibase 10.1021/acs.jpcc.5b02096}
  {\bibfield  {journal} {\bibinfo  {journal} {The Journal of Physical Chemistry
  C}\ }\textbf {\bibinfo {volume} {119}},\ \bibinfo {pages} {6918--6922}
  (\bibinfo {year} {2015})}\BibitemShut {NoStop}%
\bibitem [{\citenamefont {Kane}\ and\ \citenamefont {Mele}(2005)}]{z2kane}%
  \BibitemOpen
  \bibfield  {author} {\bibinfo {author} {\bibfnamefont {C.~L.}\ \bibnamefont
  {Kane}}\ and\ \bibinfo {author} {\bibfnamefont {E.~J.}\ \bibnamefont
  {Mele}},\ }\bibfield  {title} {\enquote {\bibinfo {title} {${Z}_{2}$
  topological order and the quantum spin hall effect},}\ }\href {\doibase
  10.1103/PhysRevLett.95.146802} {\bibfield  {journal} {\bibinfo  {journal}
  {Phys. Rev. Lett.}\ }\textbf {\bibinfo {volume} {95}},\ \bibinfo {pages}
  {146802} (\bibinfo {year} {2005})}\BibitemShut {NoStop}%
\bibitem [{\citenamefont {Roy}(2009{\natexlab{a}})}]{z2Rahul2}%
  \BibitemOpen
  \bibfield  {author} {\bibinfo {author} {\bibfnamefont {Rahul}\ \bibnamefont
  {Roy}},\ }\bibfield  {title} {\enquote {\bibinfo {title} {${Z}_{2}$
  classification of quantum spin hall systems: An approach using time-reversal
  invariance},}\ }\href {\doibase 10.1103/PhysRevB.79.195321} {\bibfield
  {journal} {\bibinfo  {journal} {Phys. Rev. B}\ }\textbf {\bibinfo {volume}
  {79}},\ \bibinfo {pages} {195321} (\bibinfo {year}
  {2009}{\natexlab{a}})}\BibitemShut {NoStop}%
\bibitem [{\citenamefont {Roy}(2009{\natexlab{b}})}]{z2Rahul}%
  \BibitemOpen
  \bibfield  {author} {\bibinfo {author} {\bibfnamefont {Rahul}\ \bibnamefont
  {Roy}},\ }\bibfield  {title} {\enquote {\bibinfo {title} {Topological phases
  and the quantum spin hall effect in three dimensions},}\ }\href {\doibase
  10.1103/PhysRevB.79.195322} {\bibfield  {journal} {\bibinfo  {journal} {Phys.
  Rev. B}\ }\textbf {\bibinfo {volume} {79}},\ \bibinfo {pages} {195322}
  (\bibinfo {year} {2009}{\natexlab{b}})}\BibitemShut {NoStop}%
\bibitem [{\citenamefont {Giannozzi}\ \emph {et~al.}(2009)\citenamefont
  {Giannozzi}, \citenamefont {Baroni}, \citenamefont {Bonini}, \citenamefont
  {Calandra} \emph {et~al.}}]{QE}%
  \BibitemOpen
  \bibfield  {author} {\bibinfo {author} {\bibfnamefont {Paolo}\ \bibnamefont
  {Giannozzi}}, \bibinfo {author} {\bibfnamefont {Stefano}\ \bibnamefont
  {Baroni}}, \bibinfo {author} {\bibfnamefont {Nicola}\ \bibnamefont {Bonini}},
  \bibinfo {author} {\bibfnamefont {Matteo}\ \bibnamefont {Calandra}},  \emph
  {et~al.},\ }\bibfield  {title} {\enquote {\bibinfo {title} {Quantum espresso:
  a modular and open-source software project for quantum simulations of
  materials},}\ }\href {http://stacks.iop.org/0953-8984/21/i=39/a=395502}
  {\bibfield  {journal} {\bibinfo  {journal} {J. Phys.: Condens. Matter}\
  }\textbf {\bibinfo {volume} {21}},\ \bibinfo {pages} {395502} (\bibinfo
  {year} {2009})}\BibitemShut {NoStop}%
\bibitem [{\citenamefont {Perdew}\ \emph {et~al.}(1996)\citenamefont {Perdew},
  \citenamefont {Burke},\ and\ \citenamefont {Ernzerhof}}]{PBE}%
  \BibitemOpen
  \bibfield  {author} {\bibinfo {author} {\bibfnamefont {John~P.}\ \bibnamefont
  {Perdew}}, \bibinfo {author} {\bibfnamefont {Kieron}\ \bibnamefont {Burke}},
  \ and\ \bibinfo {author} {\bibfnamefont {Matthias}\ \bibnamefont
  {Ernzerhof}},\ }\bibfield  {title} {\enquote {\bibinfo {title} {Generalized
  gradient approximation made simple},}\ }\href {\doibase
  10.1103/PhysRevLett.77.3865} {\bibfield  {journal} {\bibinfo  {journal}
  {Phys. Rev. Lett.}\ }\textbf {\bibinfo {volume} {77}},\ \bibinfo {pages}
  {3865--3868} (\bibinfo {year} {1996})}\BibitemShut {NoStop}%
\bibitem [{\citenamefont {Hedin}(1965)}]{Hedin}%
  \BibitemOpen
  \bibfield  {author} {\bibinfo {author} {\bibfnamefont {Lars}\ \bibnamefont
  {Hedin}},\ }\bibfield  {title} {\enquote {\bibinfo {title} {New method for
  calculating the one-particle green's function with application to the
  electron-gas problem},}\ }\href {\doibase 10.1103/PhysRev.139.A796}
  {\bibfield  {journal} {\bibinfo  {journal} {Phys. Rev.}\ }\textbf {\bibinfo
  {volume} {139}},\ \bibinfo {pages} {A796--A823} (\bibinfo {year}
  {1965})}\BibitemShut {NoStop}%
\bibitem [{\citenamefont {Shishkin}\ and\ \citenamefont
  {Kresse}(2006)}]{Kresse}%
  \BibitemOpen
  \bibfield  {author} {\bibinfo {author} {\bibfnamefont {M.}~\bibnamefont
  {Shishkin}}\ and\ \bibinfo {author} {\bibfnamefont {G.}~\bibnamefont
  {Kresse}},\ }\bibfield  {title} {\enquote {\bibinfo {title} {Implementation
  and performance of the frequency-dependent $gw$ method within the paw
  framework},}\ }\href {\doibase 10.1103/PhysRevB.74.035101} {\bibfield
  {journal} {\bibinfo  {journal} {Phys. Rev. B}\ }\textbf {\bibinfo {volume}
  {74}},\ \bibinfo {pages} {035101} (\bibinfo {year} {2006})}\BibitemShut
  {NoStop}%
\bibitem [{\citenamefont {Rudenko}\ and\ \citenamefont
  {Katsnelson}(2014)}]{Rudenko}%
  \BibitemOpen
  \bibfield  {author} {\bibinfo {author} {\bibfnamefont {A.~N.}\ \bibnamefont
  {Rudenko}}\ and\ \bibinfo {author} {\bibfnamefont {M.~I.}\ \bibnamefont
  {Katsnelson}},\ }\bibfield  {title} {\enquote {\bibinfo {title}
  {Quasiparticle band structure and tight-binding model for single- and bilayer
  black phosphorus},}\ }\href {\doibase 10.1103/PhysRevB.89.201408} {\bibfield
  {journal} {\bibinfo  {journal} {Phys. Rev. B}\ }\textbf {\bibinfo {volume}
  {89}},\ \bibinfo {pages} {201408} (\bibinfo {year} {2014})}\BibitemShut
  {NoStop}%
\bibitem [{\citenamefont {Car}\ and\ \citenamefont {Parrinello}(1985)}]{Car85}%
  \BibitemOpen
  \bibfield  {author} {\bibinfo {author} {\bibfnamefont {R.}~\bibnamefont
  {Car}}\ and\ \bibinfo {author} {\bibfnamefont {M.}~\bibnamefont
  {Parrinello}},\ }\bibfield  {title} {\enquote {\bibinfo {title} {Unified
  approach for molecular dynamics and density-functional theory},}\ }\href
  {\doibase 10.1103/PhysRevLett.55.2471} {\bibfield  {journal} {\bibinfo
  {journal} {Phys. Rev. Lett.}\ }\textbf {\bibinfo {volume} {55}},\ \bibinfo
  {pages} {2471--2474} (\bibinfo {year} {1985})}\BibitemShut {NoStop}%
\bibitem [{\citenamefont {Liu}\ \emph {et~al.}(2015)\citenamefont {Liu},
  \citenamefont {Zhang}, \citenamefont {Abdalla}, \citenamefont {Fazzio},\ and\
  \citenamefont {Zunger}}]{Qihang}%
  \BibitemOpen
  \bibfield  {author} {\bibinfo {author} {\bibfnamefont {Qihang}\ \bibnamefont
  {Liu}}, \bibinfo {author} {\bibfnamefont {Xiuwen}\ \bibnamefont {Zhang}},
  \bibinfo {author} {\bibfnamefont {L.~B.}\ \bibnamefont {Abdalla}}, \bibinfo
  {author} {\bibfnamefont {Adalberto}\ \bibnamefont {Fazzio}}, \ and\ \bibinfo
  {author} {\bibfnamefont {Alex}\ \bibnamefont {Zunger}},\ }\bibfield  {title}
  {\enquote {\bibinfo {title} {Switching a normal insulator into a topological
  insulator via electric field with application to phosphorene},}\ }\href
  {\doibase 10.1021/nl5043769} {\bibfield  {journal} {\bibinfo  {journal} {Nano
  Letters}\ }\textbf {\bibinfo {volume} {15}},\ \bibinfo {pages} {1222--1228}
  (\bibinfo {year} {2015})}\BibitemShut {NoStop}%
\bibitem [{\citenamefont {Soluyanov}\ and\ \citenamefont
  {Vanderbilt}(2011)}]{PhysRevB.83.235401}%
  \BibitemOpen
  \bibfield  {author} {\bibinfo {author} {\bibfnamefont {Alexey~A.}\
  \bibnamefont {Soluyanov}}\ and\ \bibinfo {author} {\bibfnamefont {David}\
  \bibnamefont {Vanderbilt}},\ }\bibfield  {title} {\enquote {\bibinfo {title}
  {Computing topological invariants without inversion symmetry},}\ }\href
  {\doibase 10.1103/PhysRevB.83.235401} {\bibfield  {journal} {\bibinfo
  {journal} {Phys. Rev. B}\ }\textbf {\bibinfo {volume} {83}},\ \bibinfo
  {pages} {235401} (\bibinfo {year} {2011})}\BibitemShut {NoStop}%
\bibitem [{\citenamefont {Gresch}\ \emph {et~al.}()\citenamefont {Gresch},
  \citenamefont {Soluyanov}, \citenamefont {Aut\'{e}s}, \citenamefont {Yazyev},
  \citenamefont {Bernevig}, \citenamefont {Vanderbilt},\ and\ \citenamefont
  {Troyer}}]{Z2Pack}%
  \BibitemOpen
  \bibfield  {author} {\bibinfo {author} {\bibfnamefont {D.}~\bibnamefont
  {Gresch}}, \bibinfo {author} {\bibfnamefont {A.~A.}\ \bibnamefont
  {Soluyanov}}, \bibinfo {author} {\bibfnamefont {G.}~\bibnamefont
  {Aut\'{e}s}}, \bibinfo {author} {\bibfnamefont {O.}~\bibnamefont {Yazyev}},
  \bibinfo {author} {\bibfnamefont {B.~A.}\ \bibnamefont {Bernevig}}, \bibinfo
  {author} {\bibfnamefont {D.}~\bibnamefont {Vanderbilt}}, \ and\ \bibinfo
  {author} {\bibfnamefont {M.}~\bibnamefont {Troyer}},\ }\bibfield  {title}
  {\enquote {\bibinfo {title} {{Universal Framework for identifying topological
  materials and its numerical implementation in the Z2Pack software
  package}},}\ }\href {http://z2pack.ethz.ch/} {\bibinfo  {journal} {in
  preparation}\ }\BibitemShut {NoStop}%
\bibitem [{\citenamefont {Kim}\ \emph {et~al.}(2015)\citenamefont {Kim},
  \citenamefont {Baik}, \citenamefont {Ryu}, \citenamefont {Sohn},
  \citenamefont {Park}, \citenamefont {Park}, \citenamefont {Denlinger},
  \citenamefont {Yi}, \citenamefont {Choi},\ and\ \citenamefont
  {Kim}}]{Kim723}%
  \BibitemOpen
\bibfield  {journal} {  }\bibfield  {author} {\bibinfo {author} {\bibfnamefont
  {Jimin}\ \bibnamefont {Kim}}, \bibinfo {author} {\bibfnamefont {Seung~Su}\
  \bibnamefont {Baik}}, \bibinfo {author} {\bibfnamefont {Sae~Hee}\
  \bibnamefont {Ryu}}, \bibinfo {author} {\bibfnamefont {Yeongsup}\
  \bibnamefont {Sohn}}, \bibinfo {author} {\bibfnamefont {Soohyung}\
  \bibnamefont {Park}}, \bibinfo {author} {\bibfnamefont {Byeong-Gyu}\
  \bibnamefont {Park}}, \bibinfo {author} {\bibfnamefont {Jonathan}\
  \bibnamefont {Denlinger}}, \bibinfo {author} {\bibfnamefont {Yeonjin}\
  \bibnamefont {Yi}}, \bibinfo {author} {\bibfnamefont {Hyoung~Joon}\
  \bibnamefont {Choi}}, \ and\ \bibinfo {author} {\bibfnamefont {Keun~Su}\
  \bibnamefont {Kim}},\ }\bibfield  {title} {\enquote {\bibinfo {title}
  {Observation of tunable band gap and anisotropic dirac semimetal state in
  black phosphorus},}\ }\href {\doibase 10.1126/science.aaa6486} {\bibfield
  {journal} {\bibinfo  {journal} {Science}\ }\textbf {\bibinfo {volume}
  {349}},\ \bibinfo {pages} {723--726} (\bibinfo {year} {2015})}\BibitemShut
  {NoStop}%
\bibitem [{\citenamefont {Baik}\ \emph {et~al.}(2015)\citenamefont {Baik},
  \citenamefont {Kim}, \citenamefont {Yi},\ and\ \citenamefont {Choi}}]{Seung}%
  \BibitemOpen
  \bibfield  {author} {\bibinfo {author} {\bibfnamefont {Seung~Su}\
  \bibnamefont {Baik}}, \bibinfo {author} {\bibfnamefont {Keun~Su}\
  \bibnamefont {Kim}}, \bibinfo {author} {\bibfnamefont {Yeonjin}\ \bibnamefont
  {Yi}}, \ and\ \bibinfo {author} {\bibfnamefont {Hyoung~Joon}\ \bibnamefont
  {Choi}},\ }\bibfield  {title} {\enquote {\bibinfo {title} {Emergence of
  two-dimensional massless dirac fermions, chiral pseudospins, and berry's
  phase in potassium doped few-layer black phosphorus},}\ }\href {\doibase
  10.1021/acs.nanolett.5b04106} {\bibfield  {journal} {\bibinfo  {journal}
  {Nano Letters}\ }\textbf {\bibinfo {volume} {15}},\ \bibinfo {pages}
  {7788--7793} (\bibinfo {year} {2015})}\BibitemShut {NoStop}%
\bibitem [{\citenamefont {Gui}\ \emph {et~al.}(2008)\citenamefont {Gui},
  \citenamefont {Li},\ and\ \citenamefont {Zhong}}]{zhong}%
  \BibitemOpen
  \bibfield  {author} {\bibinfo {author} {\bibfnamefont {Gui}\ \bibnamefont
  {Gui}}, \bibinfo {author} {\bibfnamefont {Jin}\ \bibnamefont {Li}}, \ and\
  \bibinfo {author} {\bibfnamefont {Jianxin}\ \bibnamefont {Zhong}},\
  }\bibfield  {title} {\enquote {\bibinfo {title} {Band structure engineering
  of graphene by strain: First-principles calculations},}\ }\href {\doibase
  10.1103/PhysRevB.78.075435} {\bibfield  {journal} {\bibinfo  {journal} {Phys.
  Rev. B}\ }\textbf {\bibinfo {volume} {78}},\ \bibinfo {pages} {075435}
  (\bibinfo {year} {2008})}\BibitemShut {NoStop}%
\bibitem [{\citenamefont {Kokalj}(1999)}]{xc1}%
  \BibitemOpen
  \bibfield  {author} {\bibinfo {author} {\bibfnamefont {Anton}\ \bibnamefont
  {Kokalj}},\ }\bibfield  {title} {\enquote {\bibinfo {title} {Xcrysden---"a
  new program for displaying crystalline structures and electron densities},}\
  }\href {\doibase 10.1016/S1093-3263(99)00028-5} {\bibfield  {journal}
  {\bibinfo  {journal} {Journal of Molecular Graphics and Modelling}\ }\textbf
  {\bibinfo {volume} {17}},\ \bibinfo {pages} {176 -- 179} (\bibinfo {year}
  {1999})}\BibitemShut {NoStop}%
\end{thebibliography}%
\end{document}